\documentclass{aastex63}

\received{June 1, 2019}
\revised{January 10, 2019}
\accepted{\today}
\submitjournal{AJ}

\shorttitle{AGN winds in Sy1}
\shortauthors{Torres-Papaqui et al.}

\begin{document}

\title{The ubiquity of AGN winds in Seyfert 1 galaxies at low redshift}

\correspondingauthor{Juan Pablo Torres-Papaqui}
\email{jp.torrespapaqui@ugto.mx}

\author[0000-0002-8009-0637]{Juan Pablo Torres-Papaqui}
\affiliation{Departamento de Astronom\'{\i}a, Universidad de Guanajuato\\
Callej\'on de Jalisco S/N, Col. Valenciana CP: 36023 Guanajuato, Gto, M\'exico}

\author{Roger Coziol}
\affiliation{Departamento de Astronom\'{\i}a, Universidad de Guanajuato\\
Callej\'on de Jalisco S/N, Col. Valenciana CP: 36023 Guanajuato, Gto, M\'exico}

\author{Fernando J. Romero-Cruz}
\affiliation{Departamento de Astronom\'{\i}a, Universidad de Guanajuato\\
Callej\'on de Jalisco S/N, Col. Valenciana CP: 36023 Guanajuato, Gto, M\'exico}

\author{Aitor C. Robleto-Or\'us}
\affiliation{Departamento de Astronom\'{\i}a, Universidad de Guanajuato\\
Callej\'on de Jalisco S/N, Col. Valenciana CP: 36023 Guanajuato, Gto, M\'exico}

\author{Gabriela Escobar-V\'asquez}
\affiliation{Departamento de Qu\'imica, Universidad de Guanajuato, \\
Cerro de la Venada SN, Pueblito de Rocha, CP: 36040 Guanajuato, Gto, M\'exico}

\author{Abd\'{\i}as Morales-Vargas}
\affiliation{Departamento de Astronom\'{\i}a, Universidad de Guanajuato\\
Callej\'on de Jalisco S/N, Col. Valenciana CP: 36023 Guanajuato, Gto, M\'exico}

\author{J. J. Trejo-Alonso}
\affiliation{Facultad de Ingenier\'ia, Universidad Aut\'onoma de Quer\'etaro\\
Cerro de las Campanas S/N, CP: 76010, Santiago de Quer\'etaro, Qro., M\'exico}

\author{Marcel Chow-Mart\'inez}
\affiliation{Instituto de Geolog\'ia y Geof\'isica, Universidad Nacional Aut\'onoma de Nicaragua\\ 
Rotonda Universitaria Rigoberto L\'opez P\'erez 150 Metros al Este, C.P. 663, Managua, Nicaragua}

\author{Karla A. Cutiva-Alvarez}
\affiliation{Departamento de Astronom\'{\i}a, Universidad de Guanajuato\\
Callej\'on de Jalisco S/N, Col. Valenciana CP: 36023 Guanajuato, Gto, M\'exico}





\begin{abstract} 

The optical spectra of 3,896 Seyfert~1 (Sy1) galaxies detected with WISE at $z < 0.4$ were analyzed for evidence of outflows. In 37\% of the Sy1s the outflows appear as broad, blue-shifted, of the [OIII]$\lambda$5007\AA\, with a mean maximum velocity V$_{max}$ $\sim$1014 km s$^{-1}$, consistent with AGN winds. For each Sy1 we deduced that black hole (BH) mass, bolometric luminosity, Eddington ratio and power-law index of the continuum, which we compared with the star formation rate (SFR) and host morphology. Having separated our sample in two spectroscopic subgroups, Sy1s with only broad Balmer lines (Sy1B) and with both narrow and broad (Sy1N) lines and distinguishing those that show as outflow (Sy1Bw and Sy1Nw), we report the following differences: 1) the BH mass is systematically higher and the power-law steeper in the Sy1B-Sy1Bw than in the Sy1N-Sy1Nw, 2)~V$_{max}$ is higher in the Sy1Bw than in the Sy1Nw, correlated in both groups with the BH mass and bolometric luminosity, 3) the Eddington ratio and SFR are higher in the Sy1 with outflows, and 4) the specific star formation rates (sSFRs) of the Sy1s are normal for their morphology and mass; typical of early-type spiral galaxies in the green valley, far from the quenched regime. From these results we conclude that in Sy1s AGN winds are triggered by higher accretion rates and probably radiatively launched, and there is no clear evidence of an effect on the star formation.

\end{abstract}



\section{Introduction} \label{S1}

The discovery of a relation between the masses of super-massive black holes and the velocity dispersion of the stars in the bulge of their host galaxies \citep{Lynden-Bell1969,Magorrian1998,Ferrarese2000,Haring2004,Gultekin2009,Graham2011} suggests that the formation process of galaxies could be influenced by the active phase of their black holes (BHs). Pushing this idea forward, \citet{Silk1998} developed, in certain detail, the concept of energetic AGN gas outflows, which, in principle, could regulate the growth in mass of the bulge of galaxies by quenching their star formation. Over the past twenty years intense efforts (theoretical and observational) were invested by the international community to detect and study such AGN outflows (or AGN winds) and better understand the effects (feedback) that they could have on their host galaxies \citep{Harrison2018}. Yet, the results are still controversial and the whole subject open to debate. 

One of the most important of these debates took place in October 2017 in Leiden, where twenty of the most active researchers in the field met to make the point on the subject.\footnote{The proceeding for the debate was published online in 27 February 2018 at https://doi.org/10.1038/s41550-018-0407-2} The consensus was that, ``...there is currently no strong direct evidence for the impact of AGN on star formation in the overall galaxy population when different approaches and selection effects are taken into account'' \citep{Harrison2017}. However, also important were the discussions why this could be so. For example, in their intervention \cite{Cresci2018} concluded that although massive outflows in luminous ``active'' galaxies seem ubiquitous \citep[see also][and references therein]{Fiore2017}, observations that they suppress star formation on a large scale are inconclusive. To date, no global ``shut down''  of star formation has been reported, with observations favoring instead local effects (either quenching or triggering of star formation), and data are usually too scarce to produce meaningful statistics. The authors also added that, in a way, a negative result connecting AGN outflows to the star formation in their hosts was not totally unexpected, considering the different timescales of the two phenomena; the AGN activity happening over a short time period ($\sim 10^8$ yrs), while, once the gas is ejected from the central region of a galaxy, its effect on the interstellar gas could be delayed for a longer interval of time. Actually, assuming long delays, it could even become difficult to distinguish quenching triggered by AGN feedback from secular quenching, namely, a natural decrease of star formation due to the limited gas reservoir of galaxies \citep{Kennicutt1992,Kennicutt1994,Bait2017}. However, and as \cite{Cresci2018} also explicitly recognized, studies over larger samples of galaxies with unbiased star formation tracers are needed, ideally during the peak of the feedback epoch, which should be at the same time the AGN and star formation activities reach their maximum \citep[that is, $1 < z < 2$;][]{Madau1998,Madau2014}.  On the other hand, one could add that, assuming delayed feedback, evidence could appear long after the maximum peak of activity, that is, in large samples at lower redshifts. 

Another problem with AGN feedback is that outflows are multiphases, that is, they appear in different wavelengths, in X-rays, in optical, in Infra-Red, and even in radio \citep{Cicone2018}, and since those bands cover different ranges in temperature and density, consistent with different regions of the host galaxies, getting a full coverage is technically demanding. Moreover, the task of integrating all these different observational aspects in one consistent view can be theoretically exacting. In their intervention \cite{Cicone2018} gave two examples. One was IC 5063, a Seyfert~2 (Sy2) where the outflow manifestations at different wavelengths seem to have similar kinematics and spatial extents, suggesting they are part of the same feedback event. The other one was Mrk 231, an ultraluminous infrared galaxy (ULIRG), which has a complex activity type (a mixture of AGN and starburst) and a perturbed morphology (due to galaxy interactions or mergers), and where no clear interrelations between the different outflow phases could be established. However, these authors were also prompt in noting that their sample was small and biased toward specific cases, favoring high luminosity AGN with star bursts or galaxies with unusually high molecular (H$_2$) contents, complicating the statistical analysis. What are badly needed, they recommended, are larger samples spanning a wider range of intrinsic population properties, AGN bolometric luminosity, BH masses, and Eddington ratios, in different galaxy hosts (with different morphology) and, ``...exploring alternate tracers of star formation that can be applied to larger range in redshift''. 

One remarkable effort to extend the multiphase analysis is the statistical study made by \cite{Fiore2017}, which was based on a compilation, from the literature, of multiwavelength observations of outflows (in 94 galaxies) detected in molecular (CO and OH), ionized (H$\beta$, [OIII], H$\alpha$ and [CII]) and X-rays. Among their most robust results, they found for all these outflow phases strong correlations between the mass outflow rate, $\dot{\rm M}_{OF}$, the kinetic power, $\dot{\rm E}_{OF}$ and the bolometric luminosity, although with different slopes (that converge at high luminosity). Another significant result was that the mass loading factor, $\dot{\rm M}_{OF}$/SFR, seemed relatively high compared to starburst galaxies, which the authors suggested could be due to quenching.  However, recognizing their sample was biased toward extreme starbursts and massive galaxies, the authors had to conclude that the connection between outflows and SFR they observed in their small selective sample might not apply to less active and massive galaxies. 

This is where the study made by \cite{Woo2016} about the prevalence of outflows in 39,000 type~2 AGNs becomes significant. Using spectra from the Sloan Digital Sky Survey (SDSS) these authors were able to detect [OIII]$\lambda5007$ outflows in as much as 43.6\% of the galaxies in their samples, a remarkable high fraction considering the low resolution of the SDSS spectra ($\sim 69$\ km s$^{-1}$). They also showed that the fraction of detected outflows goes up with the AGN luminosity and increases with the Eddington ratio, from which they concluded that, since they found no connection with the radio luminosity \citep[making a search based on FIRST;][]{Helfand2015}, the outflows in their sample were most probably radiatively launched, consistent with AGN winds.

In the present study we extend the search for evidence of outflows based on the [OIII] emission line in Sy1s, namely, a type~1 AGN with higher average luminosity than Sy2s, Log(L$_{bol}) = 45.0$ in our Sy1 sample compared to Log(L$_{bol}) = 44.4$ in Sy2s up to $z \sim 0.25$ \citep[][]{Torres-Papaqui2012}, but lower than QSOs \citep[Log(L$_{bol}) \ge 45.5$ up to z = 0.3;][]{Coziol2017}, and without evidence of star bursts (that is, no ULIRG). Our final sample is composed of 3,896 SDSS spectra of Sy1s, with high signal to noise ratio (S/N $> 10$) in the continuum and high quality MIR photometry in WISE. From the SDSS spectra we extracted information about the outflow velocity, V$_{max}$, and the AGN characteristics, namely, BH mass, power-law index, bolometric luminosity, and Eddington ratio. The WISE data are used, on the other hand, to estimate the intensity of star formation in their host galaxies, based on a new calibration that relates their W2-W3 colors to their SFRs. Limiting our observations to low redshifts ($z < 0.4$) we can also determine their morphology, by applying an automatic classification method that uses the SDSS photometric parameters. Putting all these data together allows us to compare the outflows with the AGN and galaxy host characteristics in a meaningful statistical way. 

The organization of the paper is the following. In Section~\ref{S2} we describe our sample of Sy1s as detected in SDSS and WISE, explain our spectral analysis method, and expound how the outflows were detected and quantified. In Section~\ref{S3} we explain how we determined the main characteristics of the BHs and also describe how the parameters characterizing their host galaxies, SFR and morphology, were estimated. In Section~\ref{S4} we discuss the results of our statistical analysis. Our conclusions can be found in Section~\ref{S5}. In our study all the physical parameters that depend on the proper distance were calculated assuming a $\Lambda$CDM cosmology, adopting the generic parameters: $H_0 = 70$ km s$^{-1}$ Mpc$^{-1}$, $\Omega_{DM} = 0.30$, and $\Omega_{\Lambda} = 0.70$.

\section{Data analysis}\label{S2}

\subsection{Sample Selection}\label{SS2a}

Our spectroscopic sample was obtained from the Sloan Digital Sky Survey Data Release 7 \citep[SDSS DR7;][]{Abazajian2009} by cross-correlating this list with a target list of AGNs identified as Sy1 in the ``Catalog of quasars and active nuclei: 13th edition'', as compiled by \cite{VeronCetty2010}. In this this catalog, galaxies are classified as Sy1s when their spectra show prominent broad permitted lines, which is the standard definition \citep{Osterbrock2006}. Intermediate types identified as Sy1.2 and Sy1.5, where narrow  Balmer lines appear over the broad lines (in both H$\alpha$ and H$\beta$), were also included in our sample to complete the panorama of Sy1 galaxies. Other intermediate types where the broad lines are less conspicuous, like the Sy1.9 and Sy1.8, were not included. Based on their definition \citep{Rakshit2017}, there are also no narrow line Sy1s (NLSy1) in our sample. As a supplementary selection criteria we have kept only the galaxies with spectra that have a S/N~$\ge 10$ in the continuum and emission lines with S/N~$\ge 3$. Keeping only those entries with redshift $z \leq 0.4$ gave us a preliminary sample of 4,000 Sy1s. 

To retrieve the MIR data of these AGNs, we cross-correlated their positions, as found in SDSS DR7, with the positions of the entries in the AllWISE Data Release 2012\footnote{http://wise2.ipac.caltech.edu/docs/release/allsky/} \citep{Wright2010}. This was done using the CDS X-Match pipeline in VizieR \citep{Ochsenbein2000}, applying a search radius of 5$\arcsec$ around the position of each galaxy \citep[e.g.,][]{Clemens2013}. Keeping only the matches that have a contamination and confusion flag that is clear ($cc\_flags = 0$) and that have WISE fluxes with quality flags, ``ph\_qual", equal to A in all the first three bands (W1 $ = 3.353\ \mu {\rm m}$, W2 $= 4.603\ \mu {\rm m}$, W3 $= 11.561\ \mu {\rm m}$), and A or  B in the last band (W4 $= 22.088\ \mu {\rm m}$), we obtained high quality photometric data in the MIR for 3,896 Sy1s (97.4\% of our original spectroscopic sample). These Sy1s constitute our final sample. 

\subsection{Spectral analysis}\label{SS2b}

Our spectral analysis can be summarized as follows. First, we applied a correction for Galactic extinction using the extinction map of \citet{Schlegel1998} and the reddening law of \cite{Cardelli1989}. Then, we corrected for the redshift and redressed the spectra by fitting a two-component template on the continuum, containing: 1) an AGN power-law, $f_{\lambda} \propto \lambda^{\beta}$,  and 2) a Fe~{\rm II} template for the multiple iron lines. For this last correction we used  the synthetic Fe~{\rm II} template constructed by \citet{Veron2004} using high resolution spectra of {\rm I}~Zw~1, a NLSy1 with strong Fe~{\rm II} lines \citep{Vestergaard2001}. The fitting method consists in scaling and velocity broadening a Fe~{\rm II} template \citep[a method described in details in][]{Greene2005}. Note that in the Sy1s the continuum is largely dominated by the AGN (we see no stellar features), no correction due to the host galaxy was applied. After subtraction of our power-law and Fe~{\rm II} template, the average rms in the residual is 0.03 dex, which is comparable to the uncertainty introduced by this phase of the reduction as estimated by \citep{Greene2005}. 

Once the continuum was subtracted we searched for evidence of outflows. These usually appear as extended wings to the blue of the core line of [OIII], consistent with a broad component, blue-shifted by a few \AA\ \citep[e.g.,][]{Dunn2010,Sturm2011,Mullaney2013,Woo2016,Perna2017}. However, before working the oxygen line itself, we fitted and subtracted the H$\beta$ line, eliminating any possible contamination of the blue wing of the [OIII] line (in general, such contamination, once the  Fe~{\rm II} lines were eliminated, was found to be negligible). This was done by fitting Gaussian profiles to the Balmer lines that were then subtracted from the spectra.

\begin{figure*}
\epsscale{1.15}
\plotone{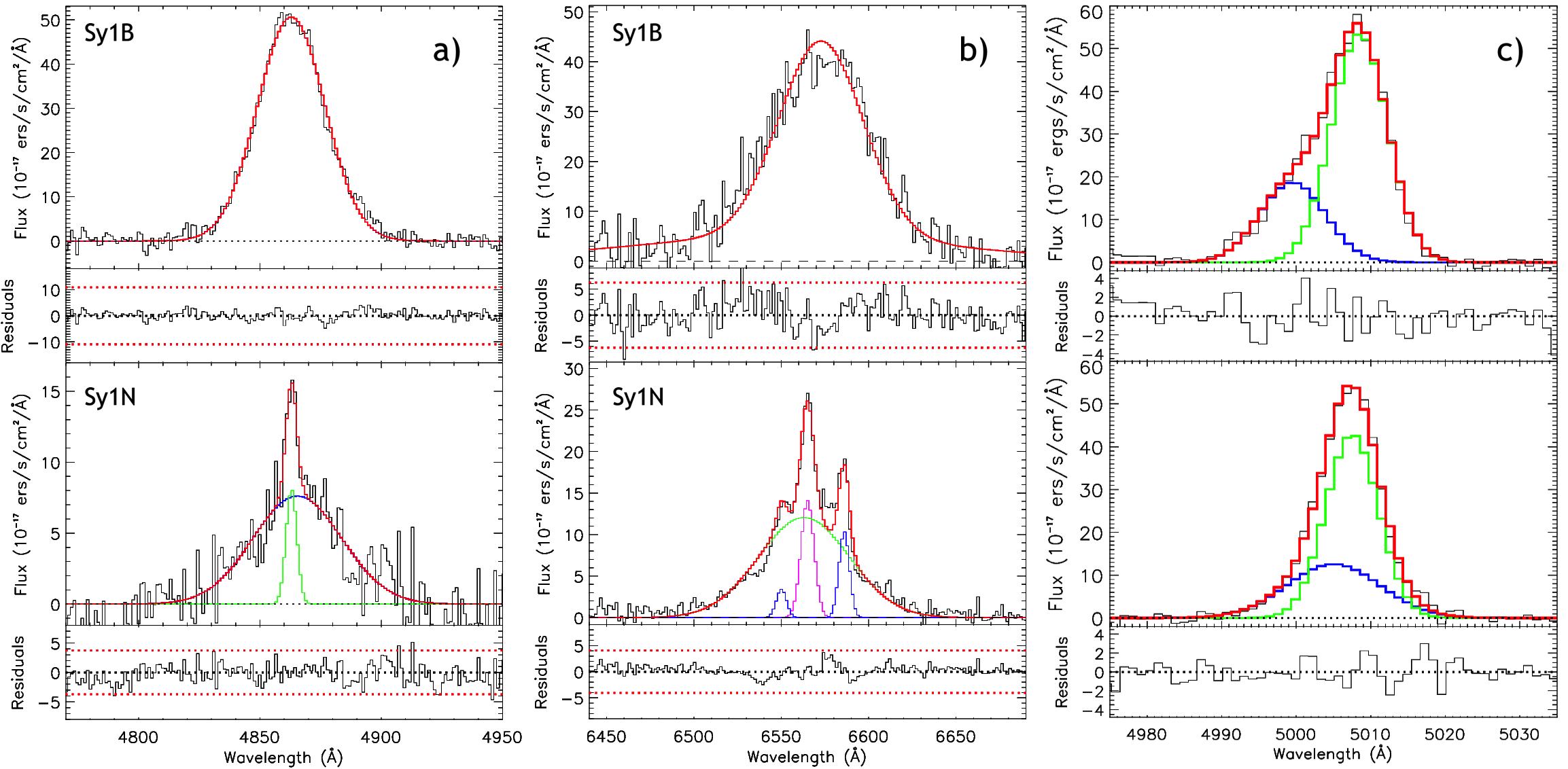}
\caption{Examples of our fitting method applied on (a) H$\beta$ and (b) H$\alpha$, distinguishing between the two Sy1 main subgroups (Sy1B and Sy1N). In each panel the different colors identify the different components and the red curve corresponds to their sum (the direct fit in the Sy1B). In (c) we show two examples of fits on [OIII]: in the upper panel the outflow is spectrally resolved, that is, its center is separated from the core by more than 69 km s$^{-1}$, which is the SDSS spectral resolution, while in the lower panel the outflow is unresolved, where the center of the outflow is separated from the core by less than 69 km s$^{-1}$.
\label{f01}}
\end{figure*}

During this part of the analysis we separated our sample of Sy1s in two: those requiring only one Gaussian were identified as Sy1B (36\% of our sample) whereas those requiring at least two Gaussians (the Sy1.2 and Sy1.5) were identified as Sy1N. Note that we did not fit outflow components to the Balmer lines, since, being limited by the low resolution of the SDSS spectra ($\sim 69$\ km s$^{-1}$), we had no way to constrain such a fit. The fit of various Gaussian profiles was done automatically by using the Levenberg-Marquardt fitting algorithm \citep{Manquardt1963} MPFIT in IDL \citep{Markwardt2009}. As initial parameters, this routine requires the central wavelengths of the line components and estimates of their amplitudes and dispersions. Iterations are then done to minimize the $\chi^2$ value in the residual, that is, after subtracting the fitted profiles from the observed lines. Two examples of Gaussian fits on H$\beta$ are shown in Figure~\ref{f01}a for the Sy1B (upper panel) and for the Sy1N (lower panel). All the fitted solutions have S/N~$\ge 3$ (the fits do not change the S/N). The fitting routine gives us the flux intensity of each component as well as its FWHM. Note that whenever there are more than one component the routine is allowed to slightly shift each Gaussian peak relative to the systemic wavelength. However, since these shifts are very small compared to the spectral resolution of the SDSS spectra, they have no measurable effect. The flux uncertainties computed on the basis of the S/N residuals are lower than 15\%, which is consistent with the mean uncertainty of the flux calibration. 

To double check our Gaussian fitting method, the same analysis was applied on H$\alpha$ (see examples in Figure~\ref{f01}b for both  Sy1B and Sy1N). For the Sy1N, this means adding two Gaussian profiles in MPFIT to fit the nitrogen doublet, [NII]$\lambda\lambda$6548,6584. Note that the most intense line of the doublet, [NII]$\lambda$6584, is almost always visible in our spectra, since the line ratio [NII]$\lambda$6584/H$\alpha$ is high in AGNs. This means that we can well constrain its position and the fact that the ratio of the intensities of the doublet must be $I_{NII6548}/I_{NII6584} = 1/3$ \citep{Osterbrock2006} allows us also to constrain the intensity of the weakest line (when too blended to be seen). 

According to \citet{Greene2005} the FWHM of the broad line components of H$\beta$ and H$\alpha$ are correlated. We verify this with our data, obtaining the relation: 
\begin{equation}\label{eq01}
{\rm FWHM}_{\rm H\beta} = (1.03\pm 0.06) \times 10^3 \left(\frac{\textrm {FWHM}_{{\textrm H}\alpha}}{10^3\, \textrm {km\, s}^{-1}}\right)^{(1.01\pm 0.03)} {\rm km}\ {\rm s}^{-1}
\end{equation}
Although our relation is comparable to the one obtained by \citet{Greene2005}, our uncertainties are consistent with a one-to-one relation, which reinforce the  conclusion that either line can be used to determine the virial mass of the SMBHs. More important for our study, however, this last result also implies that we have cleanly eliminated from the continuum the broad component in H$\beta$, suggesting that any redundant structure we detect in [OIII] must be intrinsic to this line. 

\subsection{Outflow detection}\label{SS2c}

\begin{figure*}
\gridline{\fig{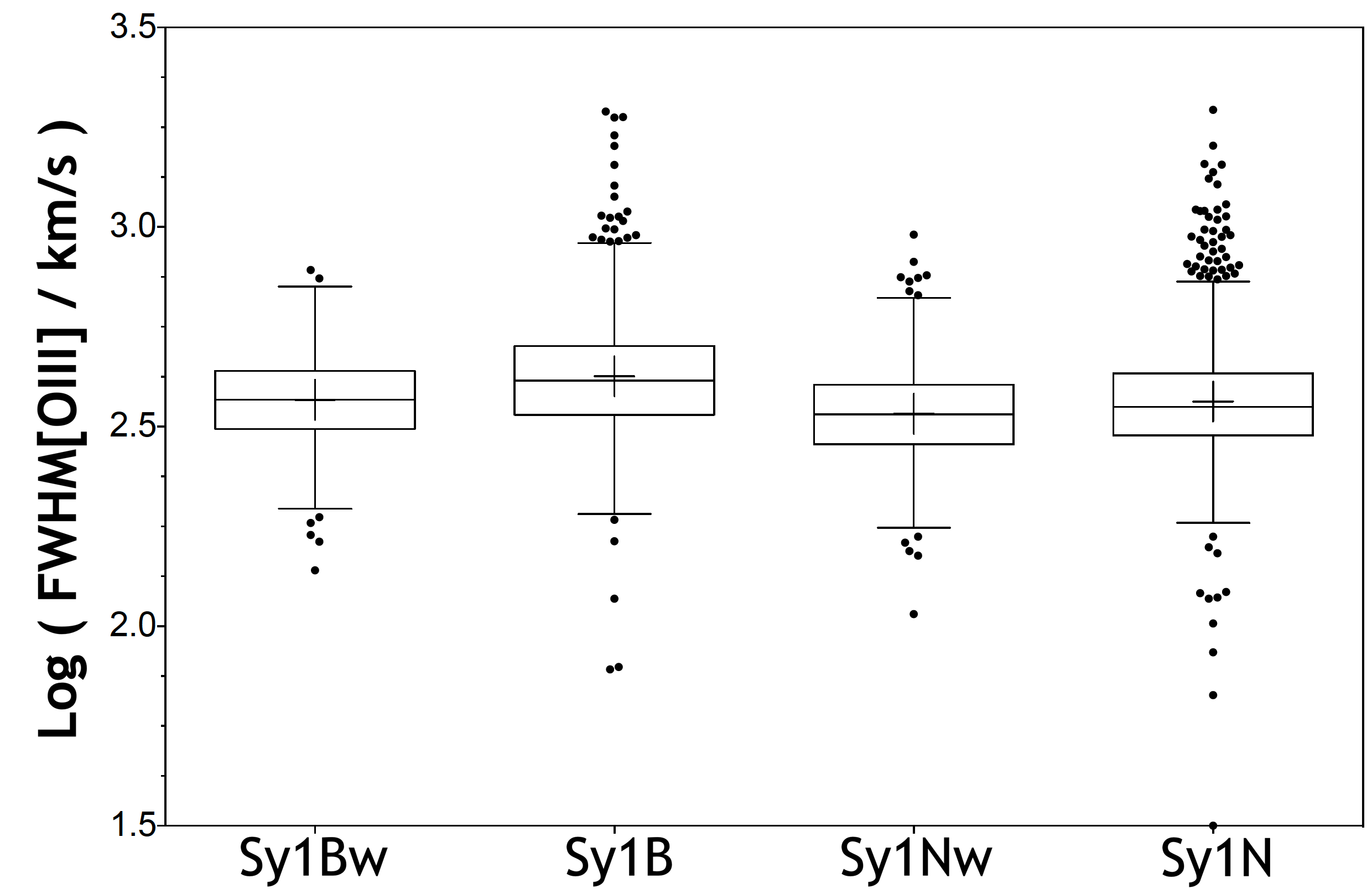}{0.56\textwidth}{(a)}
          \fig{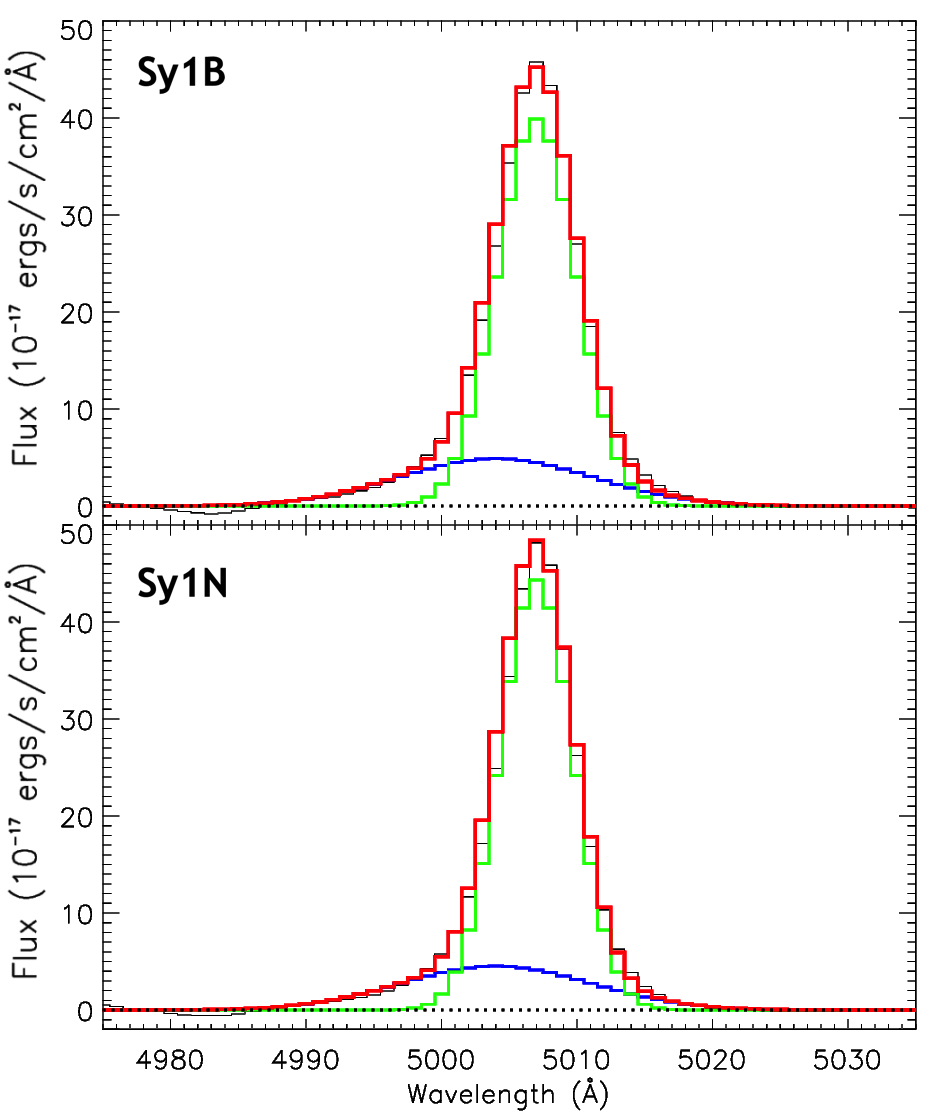}{0.3\textwidth}{(b)}
          }
\caption{(a) Tukey box-whisker plots comparing the FWHMs of the cores of the [OIII] lines in the four subgroups of Sy1s. (b) Results for the stacking of the [OIII] lines in the Sy1B and Sy1N without resolved outflows. In green we traced the line core, and in blue an broad outflow component, the sum of which is shown in red. 
\label{f2}}
\end{figure*}

In this section we now concentrate on searching for outflows in [OIII]. To quantify our search, we systematically fitted two Gaussian components: ($G$ = $G_C$ + $G_{OF}$), centering the narrow core ($G_C$) at 5007 \AA\ and fitting the outflow broad component ($G_{OF}$) as observed a few  \AA\ to the blue of this position. Note that our fitting method is limited by the resolution of the SDSS spectra, since we can only securely fit (no degeneracy) a broad component when the peak of the Gaussian is blue-shifted by more than 69 km s$^{-1}$. In Figure~\ref{f01}c in the upper panel we show one example of such resolved outflow, while in the lower panel we show one example of an unresolved solution. Taking this limit into account we found clear evidence of outflows, namely, spectrally resolved outflows, in 37\% of the total sample: 39\% in the Sy1N and 35\% in the Sy1B, identifying these cases as Sy1Nw and Sy1BNw, respectively.  Since the frequency of resolved outflows in the Sy1B and Sy1N are comparable, we can conclude that whatever causes this spectral difference, it has apparently no effect on the production of an outflow. 

From our parametric decomposition analysis, it is obvious that the low resolution of the SDSS spectra limits the detection of outflows \citep[as also noted by][]{Woo2016}. Actually, comparing in Figure~\ref{f2}a the FWHMs of the cores of the oxygen lines in the four subgroups, it seems clear that spectrally unresolved outflows must also be present, since the FWHMs are systematically larger in the Sy1B and Sy1N than in the Sy1Bw and Sy1Nw where the outflow components were subtracted. To verify the statistical significance of the differences observed, a non-parametric Kruskal-Wallis ANOVA test with Dunn's multiple comparisons test was performed.\footnote{All the statistical tests used in this article were done using GraphPad Prism version 6.00 for Mac OS X, GraphPad Software, La Jolla California USA, www.graphpad.com; A description of each test can be fond on their exhaustive guide page: https://www.graphpad.com/guides/prism/8/statistics/index.htm.} The Kruskal-Wallis test is a non-parametric test: it does not assume Gaussian distributions (three normality tests performed on our samples were rejected), but assumes their shape are similar, which is the case of the data in our samples. The p-value answers the question (the null hypothesis): if the groups are sampled from populations with identical distributions, what is the chance that random sampling would result in a sum of ranks as far apart as observed in this experiment? This implies that when the p-value is small (we choose a level of significance $\alpha = 0.05$), you can conclude that the populations have different distributions. Dunn's multiple comparisons test, then, compares the difference in the sum of ranks between two groups with the expected average difference (based on the number of groups and their size). The p-value takes into account the number of comparisons. If the null hypothesis is true (all differences between the groups are due to random sampling), then there is a 5\% chance that at least one of the post tests will have p $< 0.05$, the 5\% chance applying to the entire family of comparisons. In order to make the p-value more significant, we used the interpretation scale described in Table~\ref{STAT}. 

\begin{deluxetable*}{ccc}\label{STAT}
\tablecaption{Significance levels of statistical tests}
\tablehead{
\colhead{p-value}  & \colhead{Interpretation}                 	&\colhead{summary} 	
}
\startdata
\hspace*{1.3cm} $p < 0.0001$ &  Extremely significant  & ****	\\
$0.0001\ge p < 0.001$    	        &  Highly significant 		  &  *** 	\\
$0.001\ge p < 0.01$    	            &  Very significant 		  &  ** 	\\
$0.01\ge p < 0.05$			        &  Significant      	  	      & * 		\\
\hspace*{0.9cm} $p \ge 0.05$	&  Not significant 			  & ns 	\\
\enddata
\end{deluxetable*}

Applying the statistical test to the FWHMs of the cores of the [OIII] lines in Figure~\ref{f2}a, extremely significant differences are confirmed between the Sy1s with and without resolved outflows (the groups Sy1B-Sy1Bw and Sy1N-Sy1Nw). To put the hypothesis of unresolved outflows on a more robust observational ground, we stacked the [OIII] spectra in the Sy1B and Sy1N subgroups. The results are shown in Figure~\ref{f2}b. In each case, we were able to fit a broad component consistent with an an outflow with a S/N $\sim 60$. This result suggests that outflows are ubiquitous in Sy1s. 

The Gaussian fit for the broad component of the [OIII] line gives us two parameters that are important to qualify outflows as AGN winds \citep{Karouzos2016}: the drifting velocity, $V_{OF}$ (positive), that is, how much the Gaussian peak is shifted to the blue (after correcting for resolution), and the velocity dispersion of the line, $\sigma_{OF}$. Following \citet{Rupke2013} and \citet{Fiore2017} we define the maximum velocity of the wind as:
\begin{equation}\label{Vmax}
{\rm V}_{max} = {\rm V}_{OF} + (2 \times \sigma_{OF})\ {\rm km}\ {\rm s}^{-1}
\end{equation} 

\begin{figure*}
\gridline{\fig{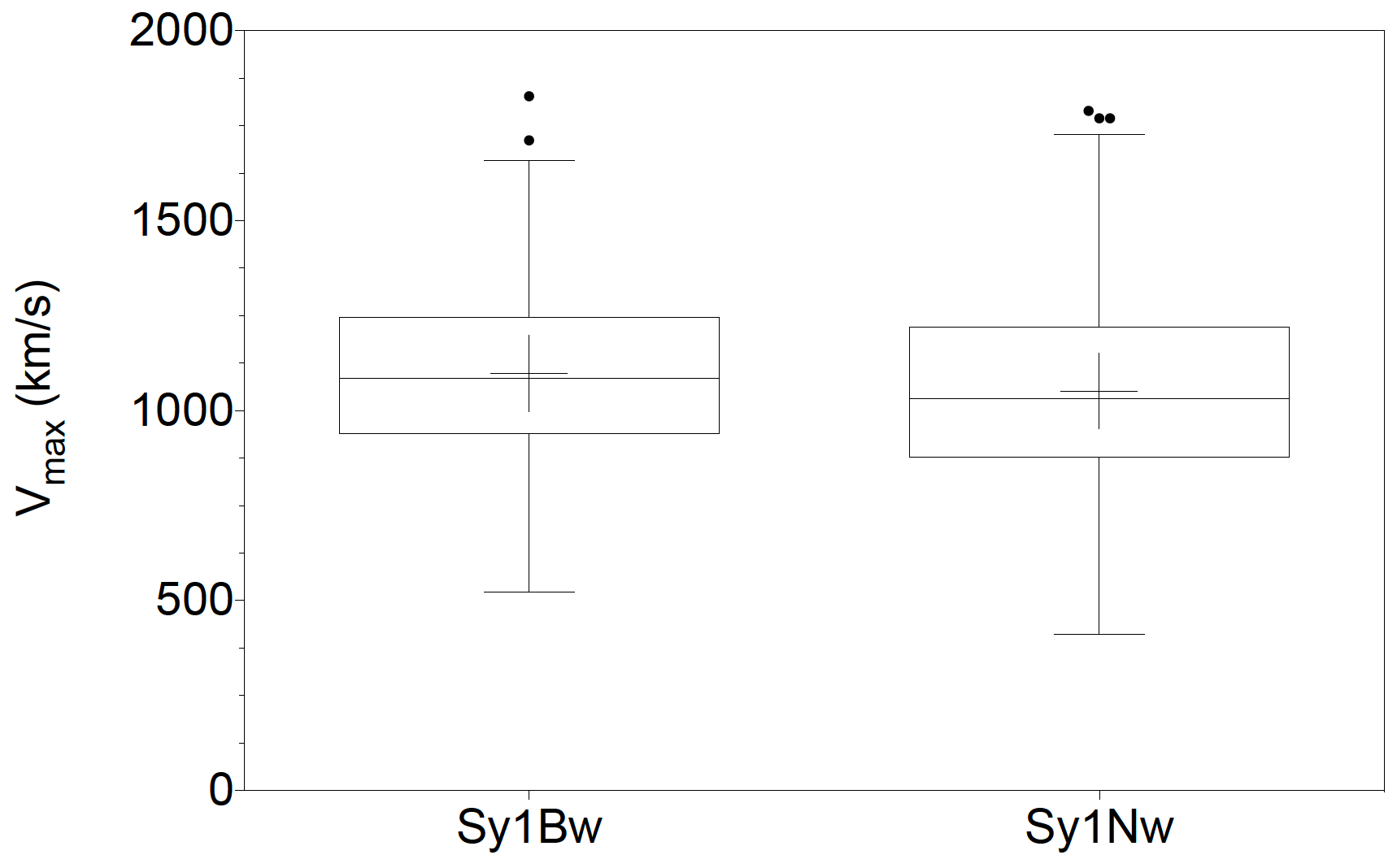}{0.5\textwidth}{(a)} 
				\fig{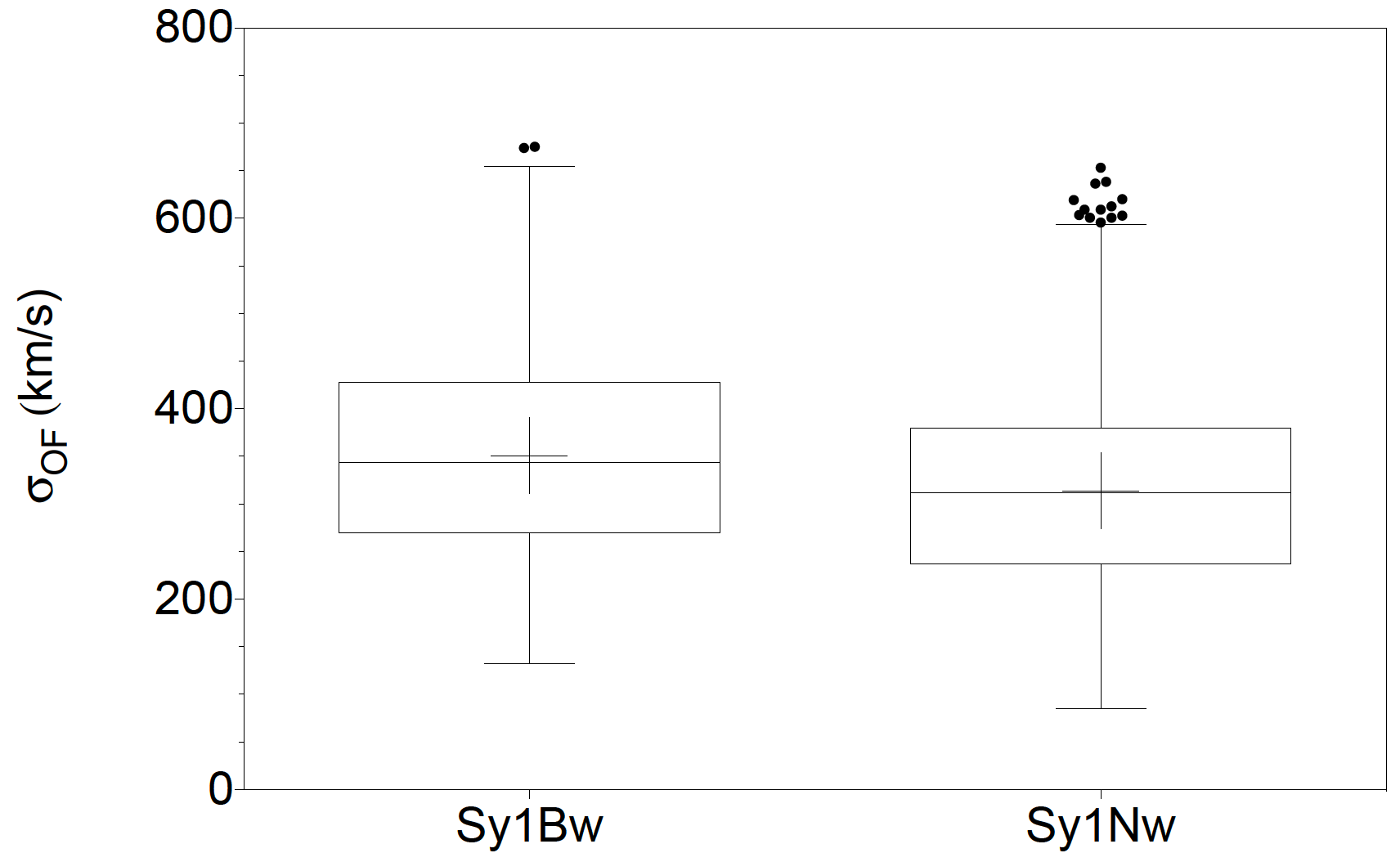}{0.5\textwidth}{(b)}
          }
\gridline{         \fig{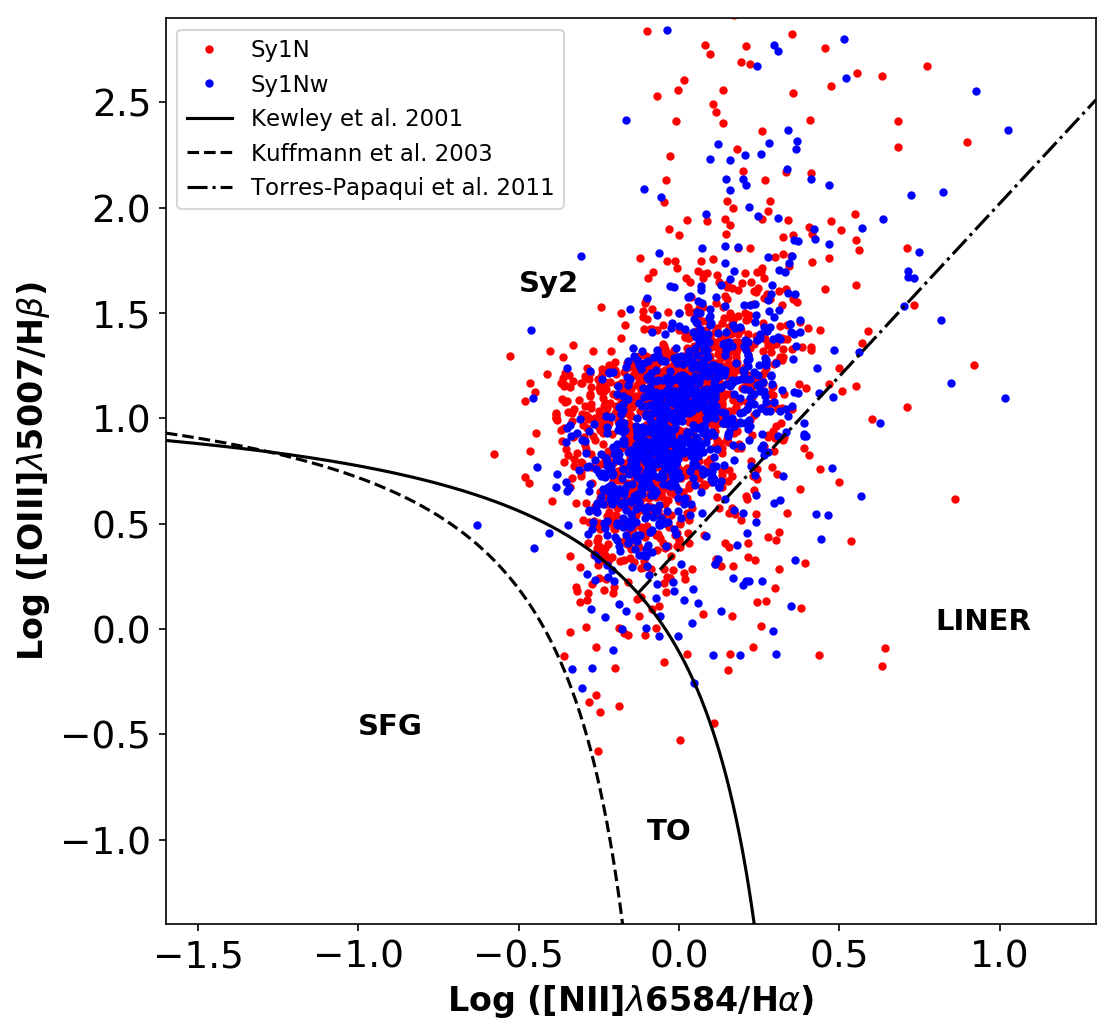}{0.5\textwidth}{(c)}
}
\caption{ (a) Tukey box-whisker plots comparing the velocity of the outflows, $V_{max}$, in the Sy1Bw and Sy1Nw. (b) Tukey box-whisker plots comparing the velocity dispersion, $\sigma_{OF}$, in the Sy1Bw and Sy1Nw. (c) BPT-VO diagnostic diagram for the Sy1Nw and Sy1N. The different regions identify galaxies with NLR excited by different sources: SFG, star forming galaxy, TO, transition type object, excited by AGNs and star formation, Sy2, excited by an AGNs, LINER, excited by low luminosity AGN. Note that a correction was applied to each of the line ratio by estimating the amount of internal extinction based on the ${\rm H}\alpha/{\rm H}\beta$ ratio and using an extinction law with  $R_V = 3.1$ \citep{Cardelli1989}.
\label{f3}}
\end{figure*}

In Figure~\ref{f3}a, we compare V$_{max}$ in the Sy1Bw and Sy1Nw. There seems to be a difference in the distributions, the Sy1Bw having slightly higher V$_{max}$ than the Sy1Nw. Performing a Mann-Whitney test, a non parametric test that is mostly sensitive to change in the median, the difference is not confirmed as significant. However, applying a Kolmogorov-Smirnov test, another non parametric test that is sensitive to any difference between the two distributions (in particular shape or spread), the difference is highly significant, with a p-value of 0.0002. Considering the definition of V$_{max}$ (Eq.~\ref{Vmax}), a difference in the velocity dispersion of the line, $\sigma_{OF}$, as shown in Figure~\ref{f3}b, explains this difference. This time a Mann-Whitney test find the difference in $\sigma_{OF}$ to be extremely significant. Finding a possible difference in $V_{max}$ is interesting, because in terms of AGN wind model it could suggest different physical conditions in the ISM of the Sy1B and Sy1N \citep{KingPounds2015}. In particular, if the Sy1N have more gas in their narrow line regions (NLRs) than the Sy1B, then the outflows could have had more difficulties to expand (because of the higher density), which would explain the lower wind velocities.

\begin{deluxetable*}{ccccccc}\label{Ap}
\tablecaption{Size of regions covered by aperture at different redshifts}
\tablehead{
\colhead{z}  & \colhead{scale}                 	&\colhead{diameter} 	& \colhead{Sy1B} 			& \colhead{Sy1Bw} 	 		& \colhead{Sy1N} 	 		& \colhead{Sy1Nw} 	\\
\colhead{ }   & \colhead{(kpc/$\arcsec$)} &\colhead{(kpc)}   		& \colhead{(\%)} 	& \colhead{(\%)} 	& \colhead{(\%)} 	& \colhead{(\%)}
}
\decimalcolnumbers
\startdata
0.1 	&  1.8   & 5.5 	& 5    & 4   & 13 	& 9 \\
0.2   	&  3.3 	&  9.9  & 30 	 & 24 & 39 	& 29 \\
0.3   &  4.4 	& 13.4 & 41 	 & 45 & 36	& 36 \\
0.4   	&  5.4 	& 16.1 & 24  & 28 & 12	   & 25 \\
\enddata
\end{deluxetable*}

Although we cannot check for a difference of NLR in the Sy1N and Sy1B, we can check to see if the presence of an outflow affects the physical state of the gas in the NLR. We do this by tracing the BPT-VO diagram \citep{Baldwin1981,Veilleux1987} in Figure~\ref{f3}c, which is used to define the level of excitation of the gas in the NLR and determine the main source of this excitation, either AGN or star formation. In Figure~\ref{f3}c  we can see that the line ratios in the Sy1N and Sy1Nw are similar to those observed in Sy2s. Taken at face value, this result is consistent with the standard AGN unification model, which states that Sy1s and Sy2s have similar engines \citep{Antonucci1985, Antonucci1993, Urry1995}. But more relevant for our analysis, the fact that we see no difference of line ratios in the Sy1N and Sy1Nw suggests that the presence of an outflow has no effect on the level of excitation of the gas in the NLR \citep[in particular, LINER-like line ratios would have been expected for shocks with AGN winds;][]{Veilleux1987}. 

At this point of our study, it seems necessary to examine what could be the importance of the aperture on our results. The aperture of the SDSS fiber has a diameter of $\sim 3\arcsec$  in the sky. As the redshift increases, the spatial regions covered by the spectra increase, and thus different parts of the galaxies are sampled: mostly the bulge at low redshifts, but larger portions of the disks as the redshift increases. To quantify this change, we compiled in Table~\ref{Ap} the projected diameters of these regions based on the cosmology used in our study.\footnote{Values calculated using Ned Wright Cosmology Calculator: http://www.astro.ucla.edu/~wright/CosmoCalc.html} From this table, we can deduce that even at low redshifts the spectra cover regions of the order of kpc, that is, extending already in the NLR. In order to see if these changes had any effect on the Sy1 subgroups (for example, favoring the detection  of Sy1N at high redshifts, or influencing the detection frequency of outflows) we calculated in the last four columns the fraction of galaxies at different redshifts in each spectral subgroup. From this table we can infer that there are no significant differences between these distributions, which suggests that the finite aperture of the fiber cannot explain the differences we observe between the Sy1 subgroups. In other words, these differences are most probably real, due to different physical characteristics.   

\section{Characterization of the AGNs and their hosts}
\label{S3}

\subsection{Determining the BH characteristics}\label{SS3a}

\begin{figure*}
\gridline{\fig{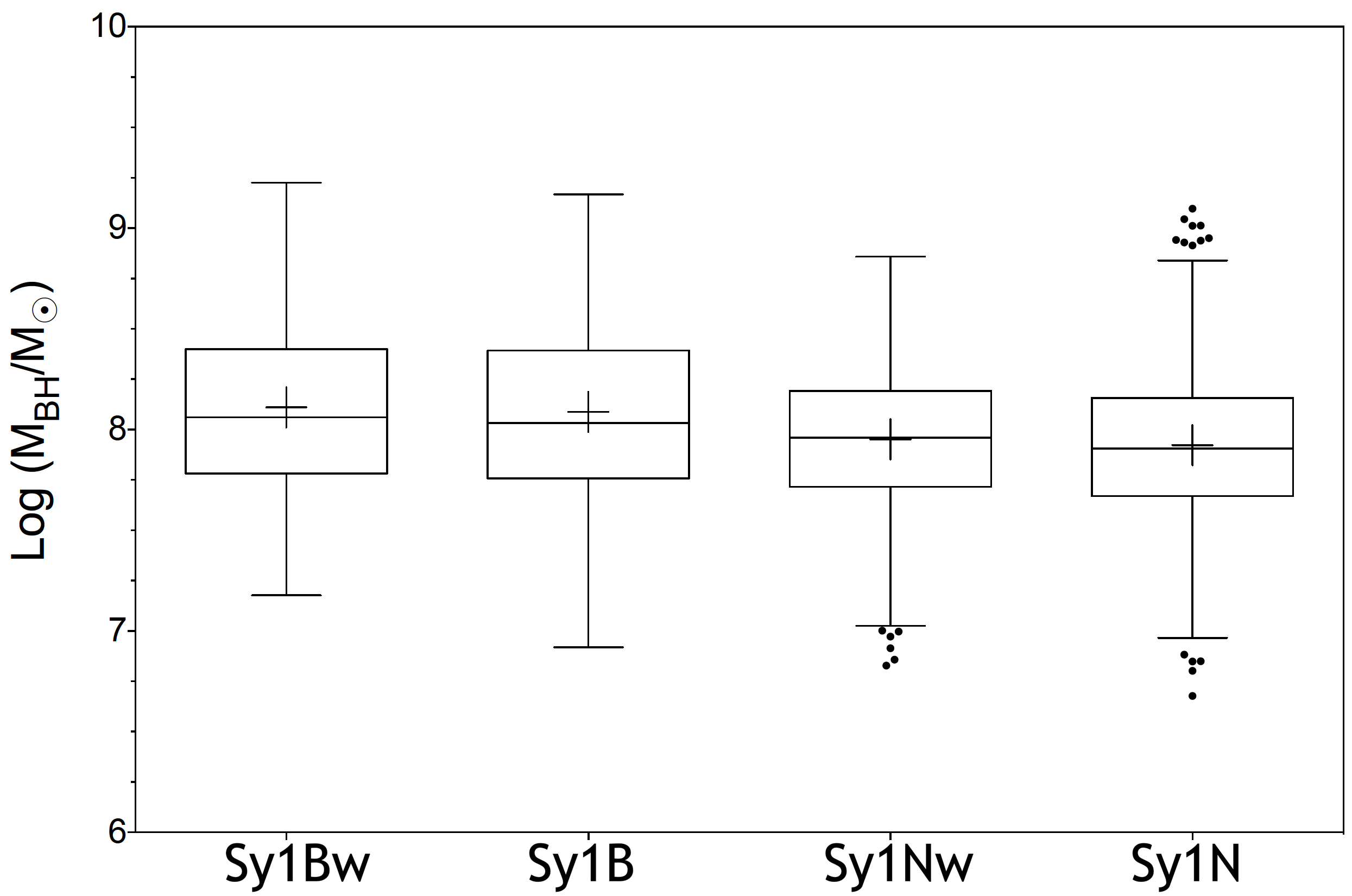}{0.35\textwidth}{(a)}
          \fig{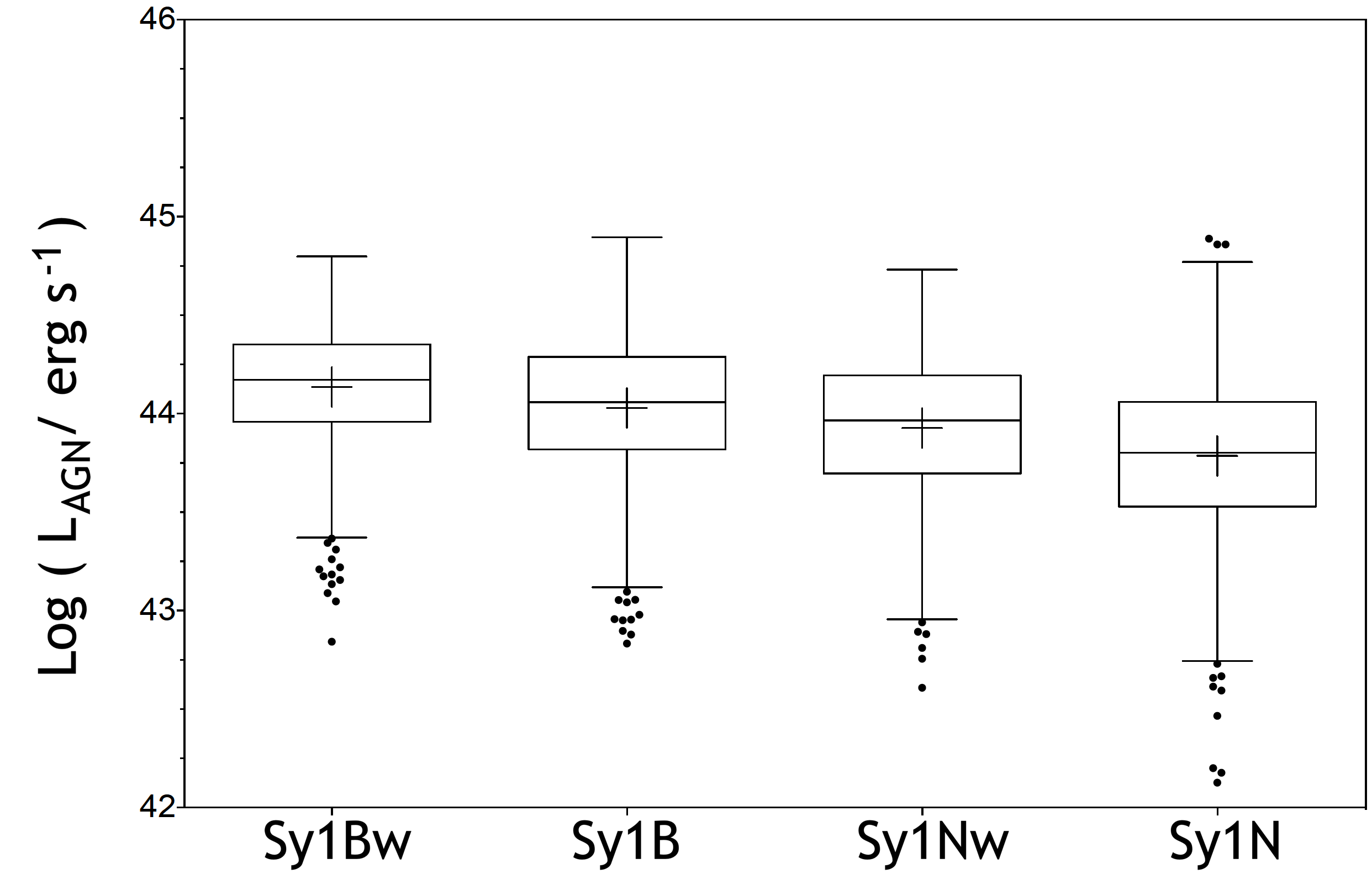}{0.35\textwidth}{(b)}
          }
\gridline{\fig{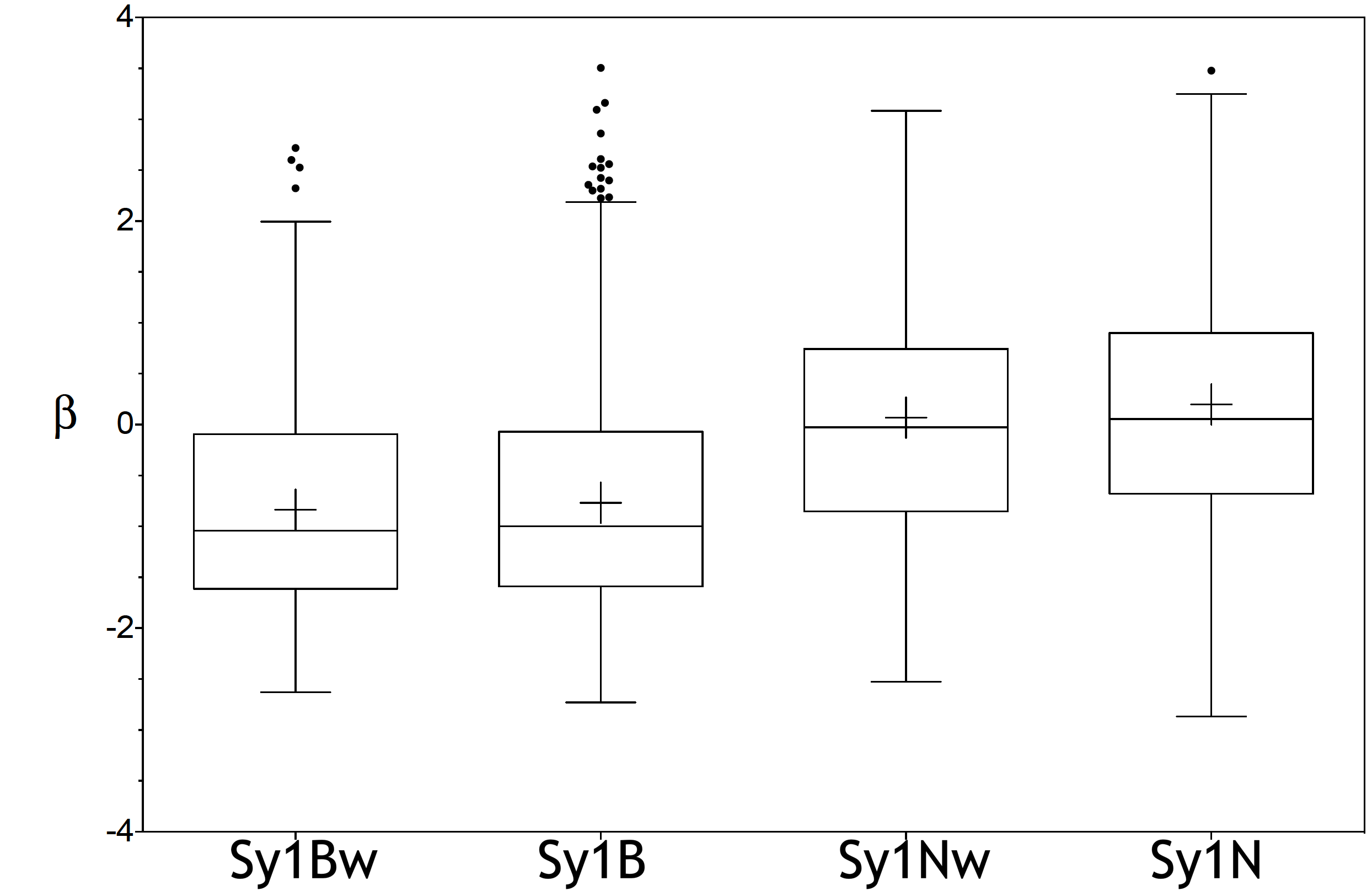}{0.35\textwidth}{(c)}
		\fig{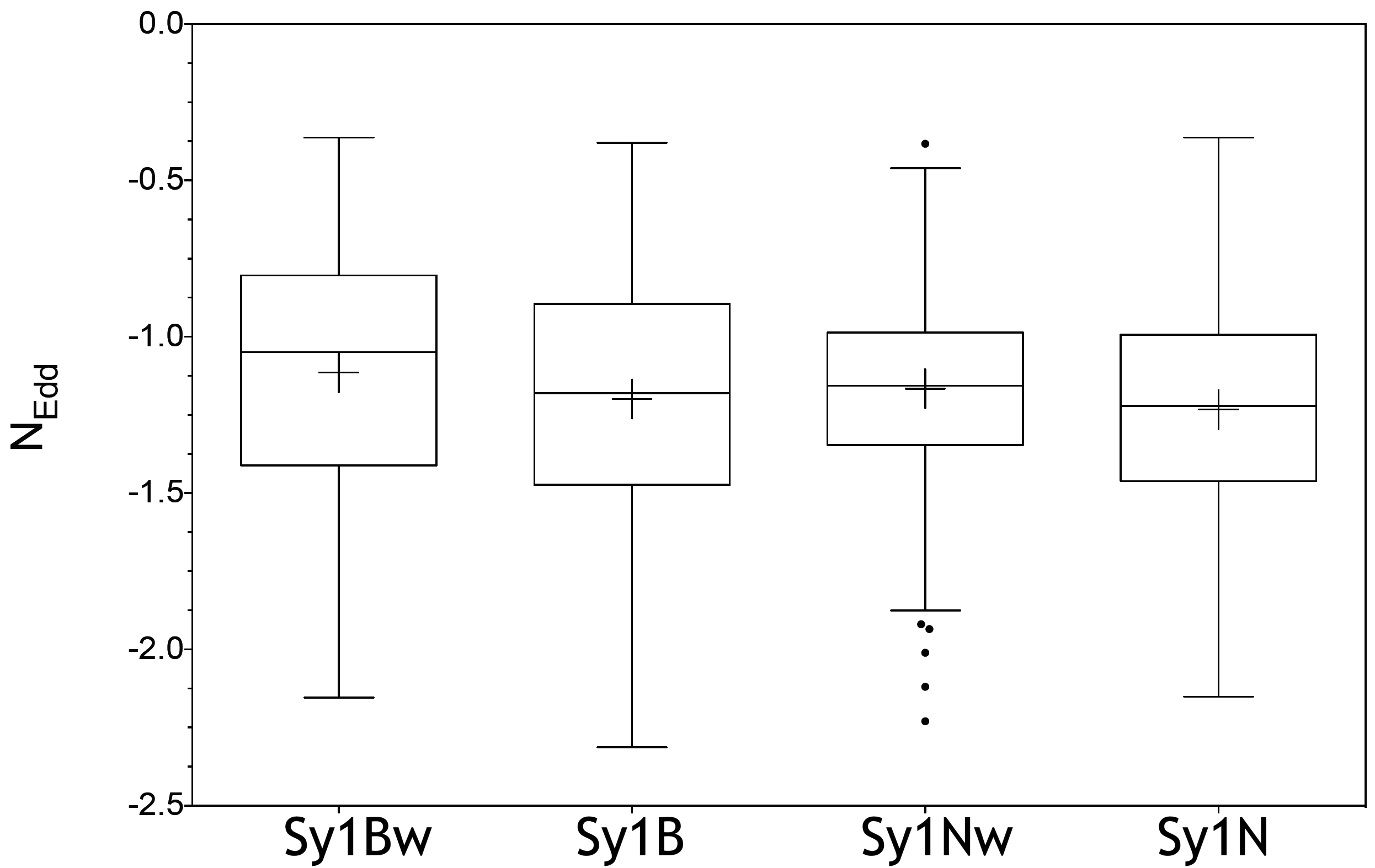}{0.35\textwidth}{(d)}
          }
 \gridline{\fig{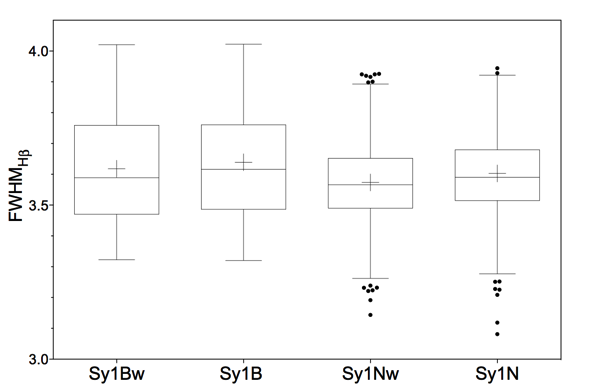}{0.4\textwidth}{(e)}
 }
\caption{Tukey box-whisker plots comparing in the four subgroups of Sy1s (a) the black hole mass, (b) the AGN luminosity, (c) the exponent of the power-law of the continuum, (d) the Eddington ratio, and (e) the FWHM of the broad component of the H$\beta$ line.
\label{f4}}
\end{figure*}

To characterize the AGNs we measured three important parameters: the luminosity of the continuum at 5100 \AA, L$_{AGN}$, the power-law index, $\beta$, and the mass of the BH, M$_{BH}$ \citep{Greene2005}: 
\begin{equation}\label{BH}
\left(\frac{{\rm M}_{BH}}{10^6\ \textrm {M}_{\odot}}\right) = (3.6\pm 0.2) \left(\frac{{\rm L}_{AGN}}{10^{44}\, \textrm {ergs\, s}^{-1}}\right)^{0.56\pm 0.02} 
\left(\frac{{\rm FWHM}_{\rm H\beta}}{10^3\, \textrm {km\, s}^{-1}}\right)^2 
\end{equation}
where FWHM$_{{\rm H}\beta}$ is the full width at half maximum of the fitted broad Gaussian in H$\beta$. Note that before using Eq.~\ref{BH}, we also applied a K-correction on the luminosity \citep{Weedman1988},  ${\rm L}_{\nu_e} = {\rm L}_{\nu_o} (\nu_o/\nu_e)^\alpha ={\rm L}_{\nu_o} (1 + z)^{-\alpha}$,  where we assumed $\alpha = -0.5$, similar to QSOs at low redshifts \citep[the majority being radio quiet;][]{Coziol2017}. From these measurements we also derived two other important parameters: the bolometric luminosity, using the relation L$_{bol} = 9.8 \times {\rm L}_{AGN}$ \citep{McLure2004} and the Eddington ratio:

\begin{equation}\label{N_Edd} 
{\rm N}_{Edd} = \log({\rm L}_{bol}/{\rm L}_{Edd})
\end{equation}
where L$_{Edd}$ is the Eddington luminosity.

In Figure~\ref{f4}a we compare M$_{BH}$ in the different Sy1 subgroups. The Sy1Bw and Sy1B have more massive black holes than the Sy1Nw and Sy1N. The mean values were reported in Table~\ref{stat1}. It is important to note that this difference is only related with the spectral groups B vs. N, independent from the outflow. The statistical significance of these differences were confirmed using a Kruskal-Wallis tests with Dunn's multiple comparisons test, as summarized in Table~\ref{stat2}. 

In Figure~\ref{f4}b we observe a continuous decline of L$_{AGN}$ along the sequence Sy1Bw:Sy1B:Sy1Nw:Sy1N. The notable difference here is the fact that in the two Sy1s  with resolved outflows (Sy1Bw and Sy1Nw) have higher luminosities than those without a resolved outflow (also true on average in Table~\ref{stat1}). These differences establish a direct connection between the outflow and the bolometric luminosity, which is related to the AGN activity. Once again, a Kruskal-Wallis tests with Dunn's multiple comparisons test confirmed that the differences are statistically significant (Table~\ref{stat2}). 

\begin{deluxetable*}{lccccccccc}
\tablecaption{Mean characteristics of the Sy1 with and without outflows. \label{stat1}}
\tablewidth{0pt}
\tablehead{
\colhead{Sy1} & \colhead{Sample} & \colhead{\% of } &\colhead{V$_{max}$} & \colhead{$\log ({\rm M}_{BH})$} & \colhead{$\log ({\rm L}_{bol})$} & \colhead{$\beta$}  & \colhead{N$_{Edd}$} & \colhead{T} & \colhead{$\log({\rm SFR})$} \\
\colhead{Subgroups} & \colhead{sizes} & \colhead{total} &\colhead{(km s$^{-1}$)}  & \colhead{(M$_\odot$)} & \colhead{(ergs s$^{-1}$)} & \colhead{} & \colhead{} & \colhead{} & \colhead{(M$_\odot$ yr$^{-1}$)}  
}
\decimalcolnumbers
\startdata
Sy1Bw 	 &  483  	&  12.4  	& 1099 		& 8.11    	& 45.1  & -0.766	  & -1.11	& 2.08	 & -0.054	  \\
Sy1B     &  905 	&  23.2 	&  \nodata 	& 8.09 		& 45.0  & -0.833	  & -1.20	& 2.08	 & -0.099	  \\
Sy1Nw  &  974   	&  25.0 	& 1052 		& 7.95    	& 44.9  &  0.068	  & -1.17	& 3.16	 & 0.004	 \\
Sy1N     & 1534 	&  39.4 	& \nodata 		& 7.92 		& 44.8  &  0.198  	  & -1.23	& 3.07	 & -0.026  \\
\enddata
\end{deluxetable*}

\begin{deluxetable*}{lccccccc}
\tablecaption{Summaries of Dunn's multiple comparisons test \label{stat2}}
\tablewidth{0pt}
\tablehead{
\colhead{Pairs} & \colhead{M$_{BH}$} & \colhead{ L$_{AGN}$} & \colhead{$\beta$}  & \colhead{N$_{Edd}$} & \colhead{T} & \colhead{SFR} & \colhead{FWHM H$\beta$}
}
\decimalcolnumbers
\startdata
Sy1Bw vs. Sy1B  		&	ns		&	**** 	&	ns		&	****	&	ns		&	****	&	*  \\
Sy1Bw vs. Sy1Nw 	&	****	&	****	&	****	&	**		&	****	&	****	&	*  \\
Sy1Bw vs. Sy1N		& ****	&	****	&	****	&	****	&	****	&	**		&	ns \\
Sy1B  	vs. Sy1Nw	&	****	&	****	&	****	&	ns		&	****	&	****	&	****\\
Sy1B  	vs. Sy1N	   	& ****	&	****	&	****	&	*		&	****	&	****	&	*  \\
Sy1Nw vs. Sy1N	 	& ns		&	****  &	*		&	****	& ns		&	***	&	****\\
\enddata
\tablecomments{Star code explained in Table~\ref{STAT}.}
\end{deluxetable*}

Considering the difference in luminosity the first results about the BH masses being comparable in Sy1s within the same spectral groups (B or N) seems less trivial than thought. This is because in the equation for the BH mass (Eq.~\ref{BH}) two parameters must balance to make the BH masses equal in the subgroups. In particular, this implies that the FWHM in the Sy1s with outflows must be smaller than the FWHM in their counterparts without an outflow. Although very weak in Figure~\ref{f4}e, the trend seems to be there. In Table~\ref{stat2}, the results reported for Dunn's post test confirm that, except for the groups Sy1Bw-Sy1N, all the differences are statistically significant, the difference between the Sy1Nw and Sy1N being more obvious than between the Sy1Bw and Sy1B. Finding such a difference was not expected, and the physical reason why this happens is unclear, since this would be related to differences in the structures of the broad line reagions (BLRs). In terms of the Virial theorem (the basis of Eq.~\ref{BH}), smaller BLRs, closer to the nucleus, would produce larger FWHMs, which suggests that outflows pushing the gas further out would produce more extended BLRs and hence smaller FWHMs. Note, however, that we do not observe evidence of outflows in the broad Balmer components (due to our low spectral resolution), and the observation of smaller FWHMs in Sy1s with outflows goes contrary to what we observed in [OIII] (where the FWHM increases with the presence of an unresolved outflow). However, the effect of outflows on low scales (below pc for the BLRs) is not necessarily expected to be similar to those observed on larger scales (pc to kpc for the NLRs). 

In Figure~\ref{f4}c we compare the power-law index, $\beta$. Likewise $M_{BH}$ the main differences are between the spectral groups, B vs N. In Table~\ref{stat2}, Dunn's post test finds no significant difference between the Sy1B and Sy1Bw, and while the difference between the Sy1N and Sy1Nw is statistically significant, it is at the lowest level. Consequently, this result confirms that the relevant physical differences are between the groups B and N: in general, the Sy1B have harder continuum, which means they emit more UV photons than the Sy1N, a characteristic that does not depend on the presence of an outflow. 

Finally, in Figure~\ref{f4}d we compare the Eddington ratios, N$_{\rm Edd}$. What is remarkable here is the fact that the Sy1s with resolved outflows (Sy1Bw and Sy1Nw) have higher ratios than those without an outflow (Sy1B and Sy1N). This particular trait of the Sy1s with resolved outflows explains why the Dunn's post test in Table~\ref{stat2} finds no significant statistical difference between the Sy1B and Sy1Nw, despite the former having higher L$_{AGN}$ than the latter (note that the difference between the Sy1B and Sy1N is also at the lowest level of statistical significance). This result directly connects the presence of outflows to higher AGN activities. It implies that the Sy1Bw and Sy1Nw have higher luminosities for a given BH mass, which could only means they have higher accretion rates ($\dot{\rm m}_{acc}$) than the Sy1B and Sy1N (since, assuming $\eta$ is the same,  L$_{AGN} = \eta\, \dot{\rm m}_{acc} c^2$). 

\subsection{Determining the star formation rates}
\label{SS3b}

\begin{figure*}
\gridline{\fig{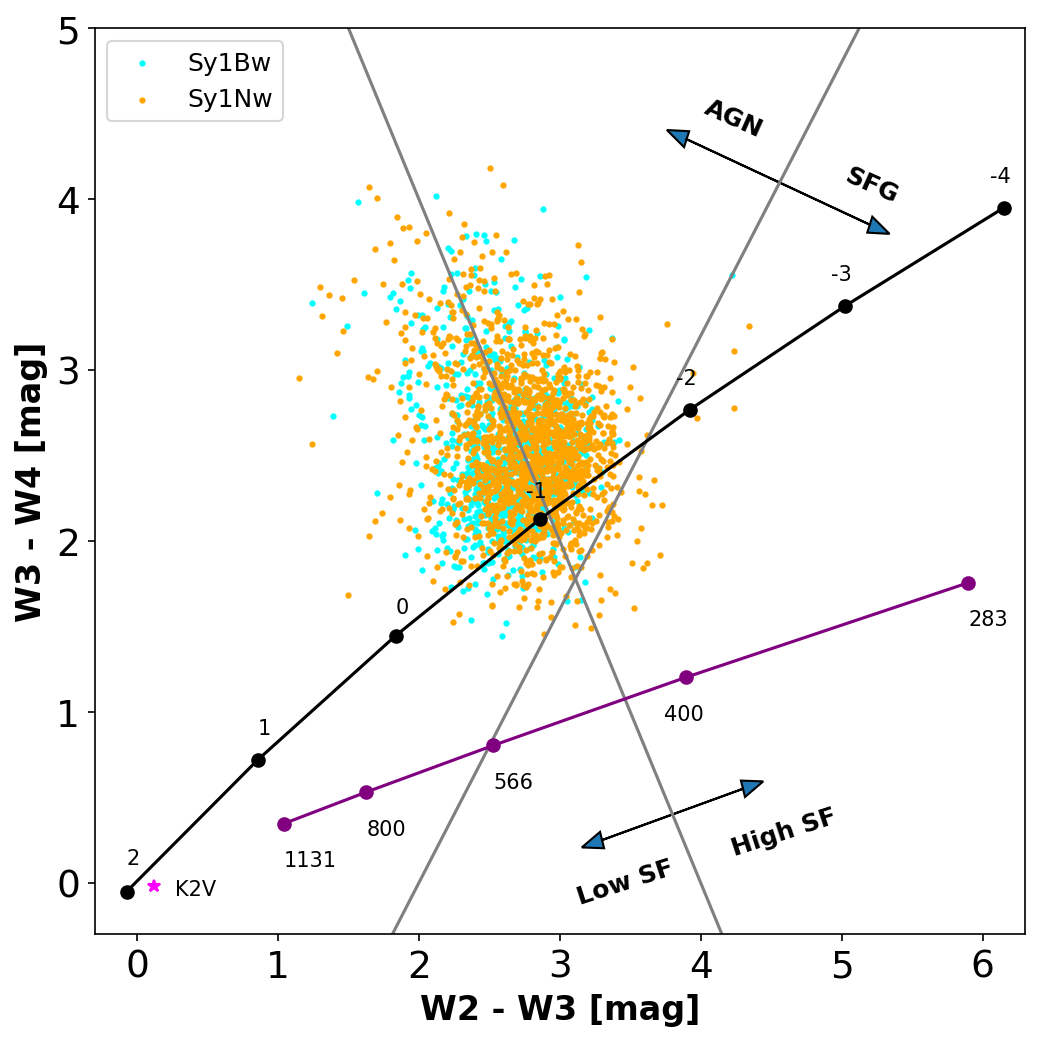}{0.45\textwidth}{(a)}
          \fig{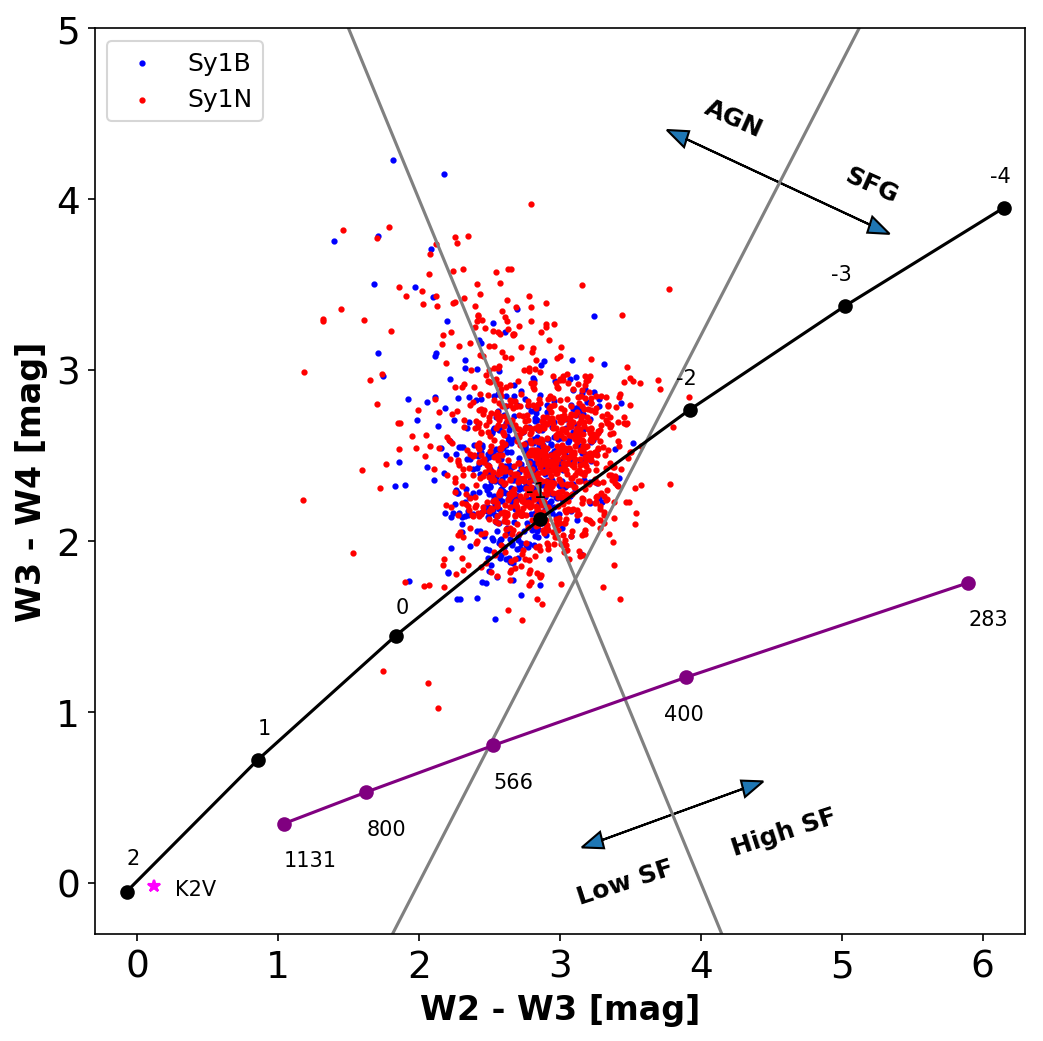}{0.45\textwidth}{(b)}
          }
\caption{MIRDD for the Sy1s distinguishing between those with resolved outflows (a), and those with unresolved outflows (b). The different regions,  lines, and values are determined as in \citet{Coziol2014}; the diagram allows to distinguish between AGN and star forming galaxies, based on low or high levels of star formation. 
\label{f5}}
\end{figure*}

The above comparisons show clear evidence of physical differences between the Sy1s with and without resolved outflows that are consistent with AGN winds: radiatively launched outflows related to higher accretion rates \citep{KingPounds2015}. Therefore, it would be natural to also expect differences in SFRs due to AGN feedback, in particular, a quenching effect of star formation in their galaxy hosts. Consequently, our main goal by computing the SFR and morphology of the AGN hosts in this section is to determine whether the SFRs of the galaxies with an outflow in our sample are somewhat peculiar relative to their morphology. 

To determine the SFR we used the W2 and W3 colors in WISE. The idea came from \citet{Coziol2014}, where a new diagnostic diagram separating AGN from SFGs, in a way similar to what can be achieved using the BPT-VO diagram, was constructed by combining the W3$-$W4 and the W2$-$W3 colors. As it was shown in this study, the working principle of this MIR diagnostic diagram (MIRDD) turned out to be the sensitivity of the W2$-$W3 color to the level of star formation in galaxies. This sensitivity was observed before and explained based on the specific MIR SEDs of star forming galaxies in \citet[][]{Jarrett2013}. A reddening of MIR colors consistent with an increase of star formation, was also found to be a common trait of type~1 AGNs at high redshifts \citep[e.g.,][]{Donoso2012,Delvecchio2014}. 

One advantage of the MIRDD over the BPT-VO diagram, is that we can compare on the same scale the level of star formation in narrow-emission line galaxies (NELGs) with the level of star formation in broad-line AGNs \citep{Coziol2015}. In Figure~\ref{f5}, we show the MIRDD for the Sy1s in each subgroup. The color distributions are almost equal. Based on the analysis done by \citet{Coziol2015} the Sy1s host galaxies would have intermediate SFRs, lower than in the Sy2s (on the high SF side) but higher than in LINERs (on the low SF side). Comparing the different Sy1s subgroups in our sample, there is a definite trend for the Sy1Bw and Sy1B to be bluer in W2$-$W3  than the Sy1Nw and Sy1N. In terms of SFR this difference is consistent with slightly higher SFRs in the Sy1Nw and Sy1N than in the Sy1Bw and Sy1B. 

Once again, we find in the case of the WISE colors a specific difference that seems to depend only on the spectral distinction, B vs. N. Therefore, could this difference  be an observational artefact, due, for example, to predominant bright MIR BHs in the B type subgroup? Comparing the SEDs of QSOs and starburst galaxies \citep[e.g.,][]{Leipski2014}, such dust hidden BHs would increase the flux in W2, reducing the difference with W3 and making the W2$-$W3 color slightly bluer (moving toward the left in the MIRDD). However, to increase the W2 flux in this way implies adding a lot of dust in the BLRs of the Sy1B and Sy1Bw, which contradicts our observation, according to the difference in $\beta$, that these galaxies produce a higher number of UV photons than the Sy1N and Sy1Nw. On the other hand, an increase in star formation in the Sy1s in the N group would increase the flux in W3, making their W2$-$W3 colors redder than those in the B group (moving toward the right in the MIRDD). Considering the possible MIR SEDs, the hypothesis of higher SFRs in the Sy1N and Sy1Nw than in the Sy1B and Sy1Bw seems much more probable. A more relevant question then is, how much higher?

\begin{figure*}
\gridline{\fig{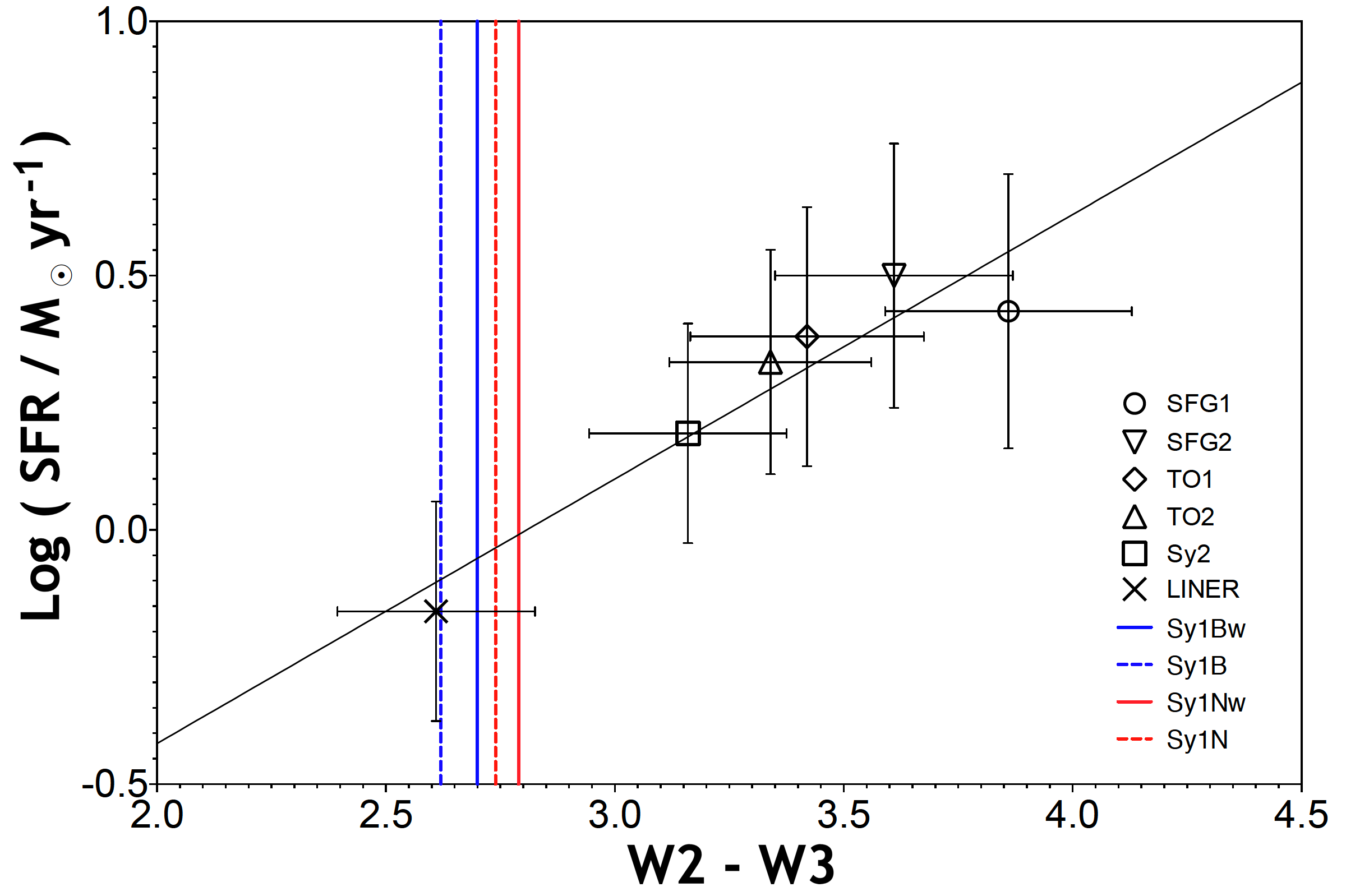}{0.49\textwidth}{(a)}
          \fig{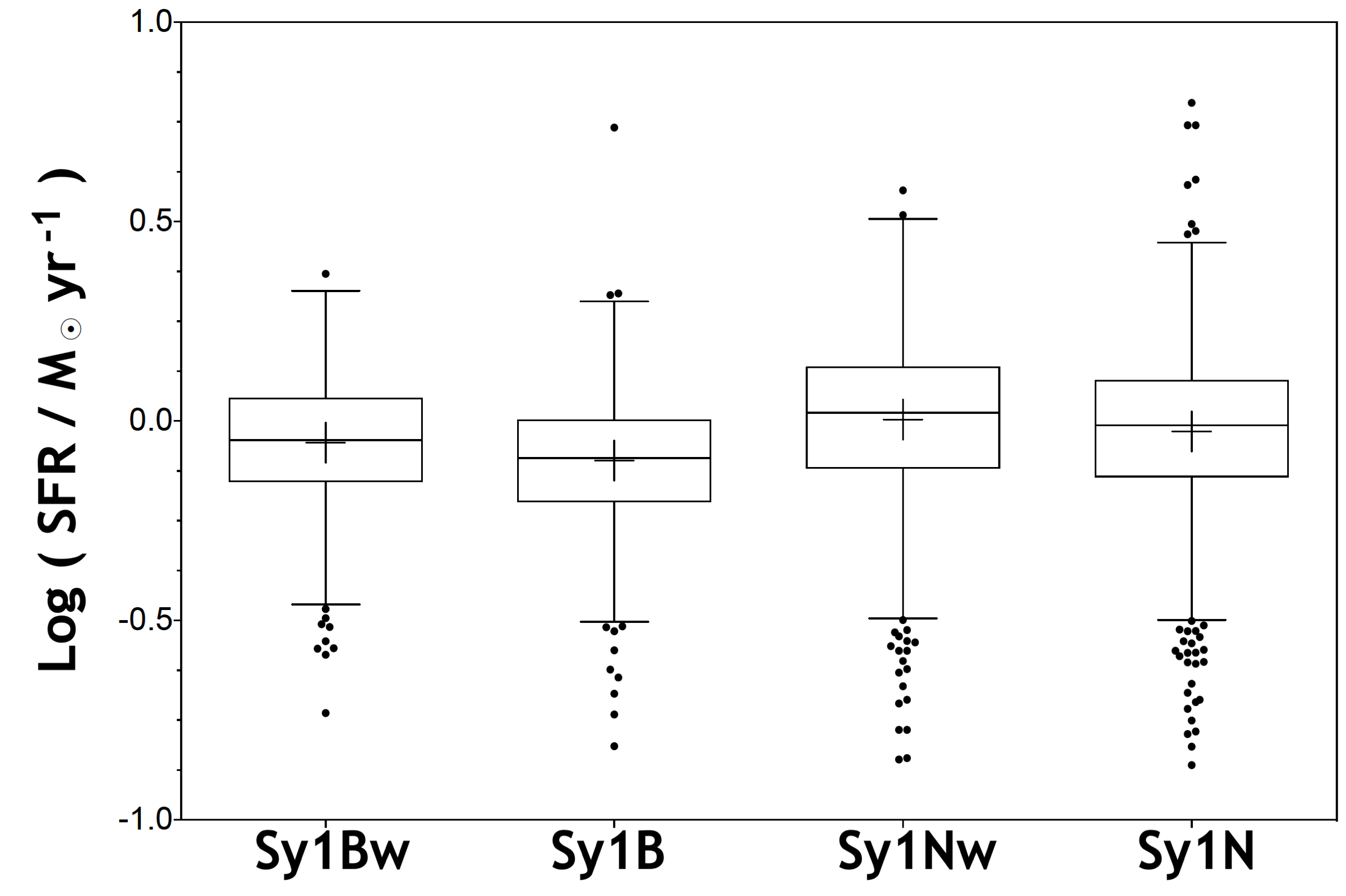}{0.48\textwidth}{(b)}
          }
\caption{SFR in the Sy1s. In (a) we show our calibration in terms of SFR of the W2-W3 colors of the NELGs; the points are the medians and the error bars the Tukey's wiskers (the farthest points where the data are not outliers). The vertical bar are the medians MIR colors in the Sy1 subgroups: dash blue, Sy1B, continuous blue Sy1Bw, dash red, Sy1N, continuous red Sy1Nw. In (b) we show the Tukey box-whisker plots comparing the SFR in the four subgroups of Sy1s. 
\label{f6}}
\end{figure*}

Using the SFRs measured in NELGs by \citet{Coziol2015}, which were established based on stellar population synthesis, we can calibrate the W2$-$W3 color in terms of SFR. In Figure~\ref{f6}a, a linear regression yields the relation: 
\begin{equation}\label{eq04}
\log\left(\frac{\textrm{SFR}}{\textrm{M}_{\odot}/\textrm{yr}}\right) =  [0.52\pm 0.03 \cdot ({\rm W2-W3})] - (1.46\pm 0.07)
\end{equation}

This linear regression has a $r^2 = 0.89$ and a p-value of 0.0044. A statistical test rejects deviations from linearity with a p-value = 0.4000. Comparing with the literature, Eq.~\ref{eq04} yields values that are in very good agreement with the SFR estimated using other methods, in optical and UV \citep[][]{Kennicutt2012}, in NIR \citep[][]{Delvecchio2014} and based on SEDs fitting \citep[][]{Bait2017}. In Figure~\ref{f6}a, our calibration suggests that the SFR increases by a factor 5 along the sequence, LINER:Sy2:TO:SFG. In Figure~\ref{f6}a, we also added vertical bars indicating the median MIR colors in the different Sy1 subgroups. These medians imply that the SFR increases along the sequence Sy1B:Sy1Bw:Sy1N:Sy1Nw, but only by small amounts. This is also confirmed in Figure~\ref{f6}b, where we compare the distributions of SFR in the subgroups. A Kruskal-Wallis ANOVA test with Dunn's multiple comparisons test confirms the statistical significance of all the differences observed (see Table~\ref{stat2}). From the mean values reported in Table~\ref{stat1} we deduce that, on average, the Sy1B, the Sy1Bw and Sy1N have SFRs that are 21\%, 13\% and 7\% lower, respectively, than in the Sy1Nw. More relevant for our study, however, the Sy1s with resolved outflows are found to have systematically higher SFRs than those without resolved OF. This last result implies that in the Sy1s with resolved outflows both the AGN and star formation activities are high, which seems inconsistent with quenching. However, a clear interpretation in terms of AGN feedback is not possible until we can compare these SFRs with the morphology of their hosts. 

\subsection{Determining the morphology}\label{SS3c}

To determine the morphology, we first tried to do it by eye by examining their SDSS images. However, this exercise seemed useless, since, although a great majority showed  extended structures, consistent with spiral galaxies, discriminating what type of spiral (early or late) seems impossible. To give you a better idea, for comparison sake, we cross-correlated the positions of the galaxies in our sample with those classified in the Galaxy Zoo project \citep{Lintott2011}. We could only recover 30\% of the galaxies in our sample, all at low redshift ($z < 0.15$). Moreover, for 75\% of the recovered Sy1Bw/Sy1B and 76\% of the S1Nw/Sy1N, the morphology reported by the observers in the Galaxy Zoo were judged unclassifiable, although they could estimate a 54\% probability for these cases to be early-type. The few that were classifiable were estimated to be spirals, with only 5\% Es (Early-type). These results convinced us to try another method based on SDSS photometry, which was developed by \citet{Shimasaku2001} and \citet{Fukugita2007}.

After downloading the photometric parameters from SDSS,\footnote{http://casjobs.sdss.org} examination of the data showed that the information for our whole sample was complete and of high quality (qualified in SDSS as detected $\ge 5$ sigma in original image). For each Sy1 in our sample we found the following photometric data: the colors $u-g$, $g-r$, $r-i$, and $i-z$ and the inverse concentration index $ C = r_{50}/r_{90}$, which is equivalent with comparing the Petrosian ratio, $R_p$, at 50\% and 90\% surface brightness ratios \citep{Blanton2001,Yasuda2001}.  As was shown by \citet{Fukugita2007}, the inverse concentration index is tightly correlated with the morphological type while the variations in SDSS colors are consistent with variations of star formations in spirals and color gradients in early-type galaxies, as reproduced by synthetic stellar model and SEDs. In \citet{Torres-Papaqui2012} we already used the photometric method to determine the morphology of NELGs up to $z \leq 0.25$, finding a high level of consistency with our stellar population synthesis study. Assuming, therefore, that the morphology of the Sy1s in our sample (and thus their SEDs) did not evolved over the time interval consistent with $z \sim 0.4$ ($\sim 4.3$\ Gyrs), applying a proper K-correction \citep[e.g.,][]{Blanton2007} would allow us to estimate the morphology of our galaxies. Our expectation, based on what we know of Sy1s at low redshifts, is that the majority should turned out to be early-type (big bulge) spirals \citep{Chen2017}. 

Our method to determine the morphology can be summarized as follows. First we correct the magnitudes for Galactic extinction and apply the K-correction determined by \citet{Blanton2007}, which was developed specially for the SDSS photometric system. Then, for each galaxy, we automatically determined weights for the five photometric parameters enumerated above, based on how close their values are from the characteristic values in galaxies with different morphology. Finally, we calculate the weighted mean of these parameters to assign a morphological classification to the galaxy on a discrete scale (identified as T), which ranges from 0 to 6 (0 = E and 6 = Irr; see Table~\ref{table_T}), with an uncertainty T $ \pm\ 0.5$.

One remaining concern in employing the photometric method for the Sy1s is about a possible influence of a bright AGN. Three possible effects could be expected: 1) at low redshifts, with the photometric aperture covering only the central region of the galaxies, the broad component in emission may affect the colors, making them appear bluer than they really are, which would skew our classification toward late-type spirals, 2) a bright AGN in the nucleus of a galaxy might also make it look more compact, this time favoring earlier morphological types, 3) as the broad components of the emission lines pass from one band to another in the red, at certain redshifts the colors could abruptly change; for example, H$\beta$ passes from the $g$ to the $r$ band around $z \sim 0.13$, making the $g-r$ redder, and H$\alpha$ passes from the $i$ to the $z$ filter around $z \sim 0.25$, making the $r-i$ redder and $i-z$ bluer. 

We did observe on average bluer $u-g$ and $g-r$  colors (by 0.3 and 0.2 mag respectively) for the Sy1B and Sy1Bw compared to the Sy1N and Sy1Nw, which might produce a bias of our classification toward late-type spirals for the former. We also observed the expected change in $r-i$ and $i-z$ colors as H$\alpha$ changes from one band to the next with redshift, but only for the Sy1B and Sy1Bw above $z \sim 0.25$. However, what the effect could be on our classification is not clear. On the other hand, these effects were observed before the K-correction, which once applied reduced significantly the variations between the subgroups. As for the compacity parameter $C$, we observed no differences between the subgroups, all the values being consistent with early-type spirals (Sa/Sb), and staying constant as the redshift increases. Note that the absence of variation of $C$ with redshift for galaxies with similar morphology was a criterion used by \citet{Shimasaku2001} and \citet{Fukugita2007} to legitimize their photometric method. Similarly, we saw no variation of the $b/a$ parameter with the redshift ($b$ and $a$ being the semi-minor and semi-major axes, respectively), which suggests that most galaxies in our sample are seen face on ($b/a \sim 0.8)$, which, for Sy1s, is also consistent with the standard AGN unification model; although one cannot assume that there is necessarily a connection between the orientation of the galaxy and the orientation of the obscuring torus; \citet{WuHan2001}. 

\begin{figure}[ht!]
\epsscale{0.9}
\plotone{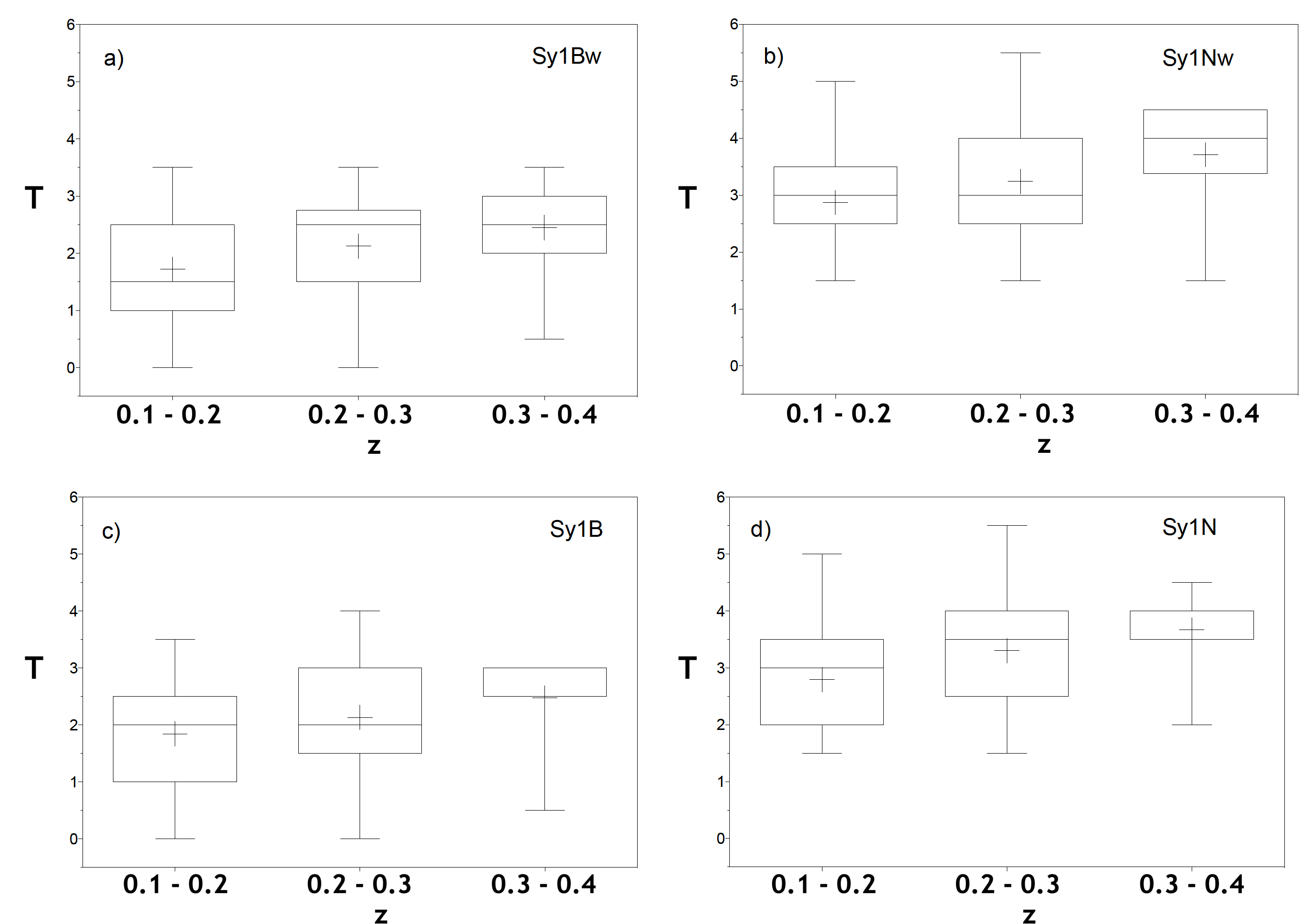}
\caption{Tukey box-whisker plots comparing the morphology among the four subgroups of Sy1s separated in three redshift bins.
\label{f07}}
\end{figure}

In Figure~\ref{f07} we compare the morphological index, T, obtained with the photometric method, separating each subgroup in three different redshift bins. These bins correspond to average spherical apertures 7.7, 11.7, and 14.8 kpc wide. Even at low redshifts, therefore, the aperture covers more than just the nucleus of the galaxies. There is a clear difference between the Sy1s B type and N type, the host of the former having slightly earlier morphological types than the latter. Note that this is a small difference on average, Sa instead of Sb in Table~\ref{stat1}, which is observed at any redshift.  The Kruskal-Wallis ANOVA test with Dunn's multiple comparisons test (Table~\ref{stat2}) confirms the differences in morphology. Once again this difference appears to depend only on the spectral group B vs. N and not on the presence of resolved outflow. Note that the preference for early-type spirals in the Sy1B and Sy1Bw goes against the bias expected based on their bluer colors. Moreover, since the differences in morphology are observed in all the different redshift bins, this cannot be due to the passage of the broad line components to different filters (which would be expected to happen only around $z \sim 0.25$ for H$\alpha$).  The only possible bias of our method could be the general increase of late-type spirals with the redshift. This implies bluer colors at high redshifts, which in the case of the Sy1s would be as expected if the AGN activity increases with the redshift. However, remembering the results shown in Table~\ref{Ap} for the spectral aperture, showing similar distributions of the Sy1 subgroups at different redshifts, and considering that the uncertainty on the morphology is T $ \pm\ 0.5$, this bias can be judged to be relatively mild, not affecting our results significantly; in particular we can still differentiate the Sy1B and Sy1N morphologically. 

\begin{deluxetable*}{cccccccccccccc}
\tablecaption{Frequencies (\%) of morphological types for the Sy1 subgoups (compared to Sy2).
\label{table_T}}
\tablewidth{0pt}
\tablehead{
\colhead{Hubble} & \colhead{E} & \colhead{E/S0} &  \colhead{S0} &\colhead{S0/Sa} &\colhead{Sa} &\colhead{Sa/Sb} &\colhead{Sb} &
\colhead{Sb/Sc} & \colhead{Sc} & \colhead{Sc/Sd} &  \colhead{Sd} &  \colhead{Sdm/Sm} &\colhead{Im} \\
\colhead{T}       					& \colhead{0.0} & \colhead{0.5} & \colhead{1.0} & \colhead{1.5} & \colhead{2.0} & \colhead{2.5} & \colhead{3.0} & \colhead{3.5} & \colhead{4.0} & \colhead{4.5} & \colhead{5.0} & \colhead{5.5} & \colhead{6.0} 
}
\startdata
Sy1Bw 	& 0.43	& 6.90	&  {\color{blue} 9.91} 	& {\color{blue} 12.50}	& {\bf 17.67} 	& {\bf 29.53} & {\bf 20.47} 	& 2.59 			&  			& 				&  		&  		&  \\
Sy1B 	& 0.58 & 5.61 & {\color{blue} 10.28} 	& {\color{blue} 13.90} 	& {\bf 17.52} 	& {\bf 28.27} & {\bf 21.26} 	& 2.34 			& 0.23 	&  			&  		&  		&  \\
Sy1Nw &  		&  		&  			& 3.51 	& 8.95 	& {\bf 18.69} 	& {\bf 19.71} 	& {\bf 17.55} 			& {\color{blue} 16.99}	& {\color{blue} 12.91} 	& 1.59 & 0.11	&  \\
Sy1N 	&  		&  		&  			& 4.04 	& {\color{blue} 12.72} 	& {\bf 18.49} 			& {\bf20.28} 	& {\bf19.31} 	& {\color{blue} 14.90} 	& 8.68 	& 1.35 & 0.22 &  \\
Sy2* 	&  		& 0.95 & 5.49 	& {\color{blue} 10.91} 	& {\color{blue} 16.51} 	& {\bf 20.48} 	& {\bf 20.64} 	& {\bf 16.92} 			& 7.84 	& 0.26 	&  		&  		&  \\
\enddata
\tablecomments{*Data for the Sy2 comes from \citet{Torres-Papaqui2013}. The three most populated bins are identified in bold, while the trends are identified in blue.}
\end{deluxetable*}

As a further test of our method, we compare in Table~\ref{table_T} our morphological classification for the Sy1s with the classification for a large sample of Sy2s, which was obtained by \citet{Torres-Papaqui2013} using the same photometric method. The three most frequent morphological types in each subgroup are marked in bold, while the trend, that is, the next two most frequent morphology bins, are marked in blue. It can be seen that the morphology of the Sy1s and Sy2s are very similar to each other. The trend for the Sy1Nw and Sy1N is for these galaxies to be slightly later-type than the Sy2s, while the trend is the contrary for the Sy1Bs and Sy1B. Once again, these differences do not depend of the presence of a detected outflow. In general therefore, the trend for the Sy1B to be slightly earlier-type than the Sy2, and even this difference decreasing in the intermediate Sy1N, are as expected based on what we know about the different Seyfert galaxies.   

Highly relevant for our study, therefore, although the differences in morphology between the Sy1s and Sy2s are minimal, the differences in SFRs are significant, of the order of 40-50\% lower in the former than in the latter. Could this suggests unusually low SFRs in the Sy1s? However, we must also consider the possibility that their hosts do not have the same mass, since the SMBHs in the Sy2s have a median mass of only $10^{7.5}$ M$_\odot$ compared to $10^8$ M$_\odot$ for the Sy1s. Assuming the galaxy hosts are roughly a 1000 times more massive \citep[e.g.,][]{Alexander2008, KingPounds2015}, this would yield specific SFR (sSFR; the star formation rate per unit of mass) of the order of $10^{-10.5}$\ yr$^{-1}$ in the Sy2s, compared to $10^{-11}$\ yr$^{-1}$ in the Sy1s. Both of these values are typical of early-type spirals in the green valley, which puts the Sy1s, and this is the most important point of our analysis, far from the quenched regime \citep[see Figure~8 in][]{Bait2017}. In other words, the SFR of the Sy1s seem to be normal considering the morphology and typical masses of their hosts. 



 \section{Discussion}\label{S4}
 
\begin{figure}[ht!]
\epsscale{0.7}
\plotone{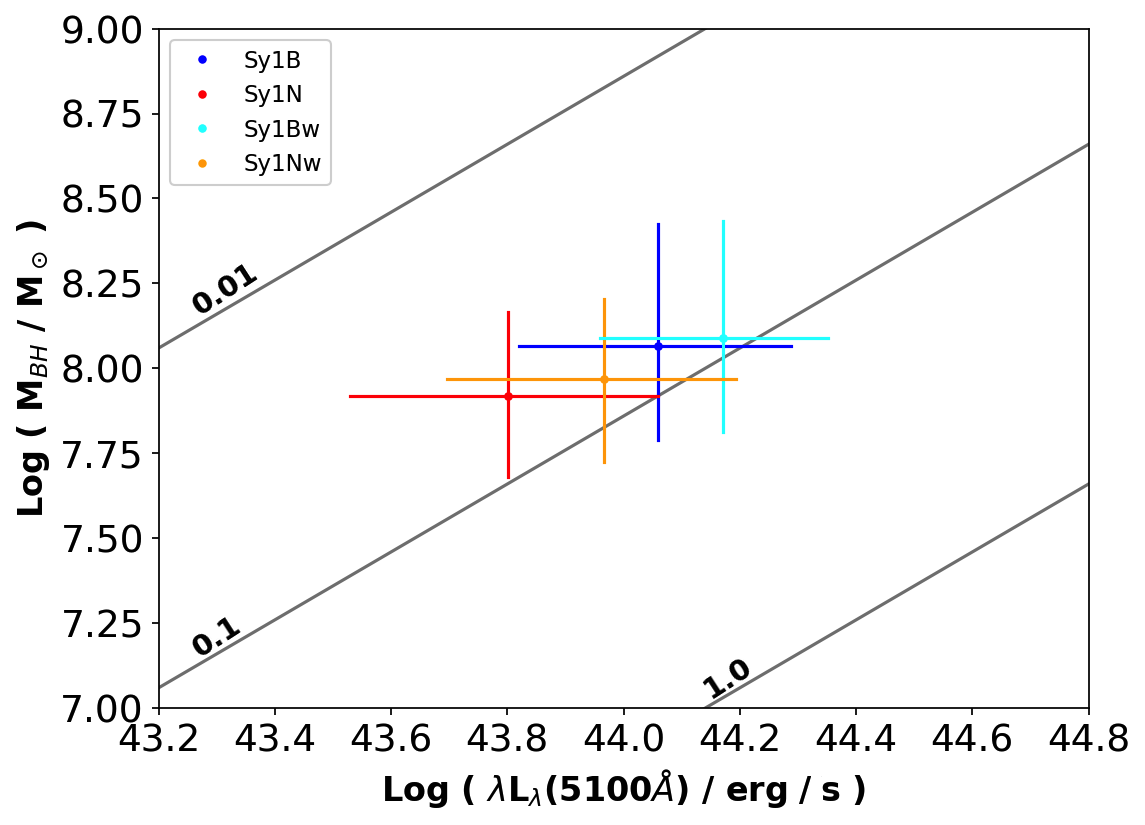}
\caption{Mass-luminosity ratios in the Sy1 subgroups. The values are the mean and standard deviations. The diagonal lines are three different Eddington ratios, N$_{Edd}$. \label{f08}}
\end{figure}

Considered as a whole, the number of Sy1s with resolved outflows at redshifts below $z \sim 0.4$ represent 37\% of our sample, which is a remarkably high fraction. Moreover, their velocities have an average value V$_{max} \sim 1014$ km s$^{-1}$ that is fully consistent with AGN winds \citep{Woo2016}. Finally, the outflows seem  connected (in Table~\ref{stat1} and Table~\ref{stat2}) to higher AGN luminosities, suggesting they could be radiatively launched. Actually, the physical reason why this could be so can be seen in Figure~\ref{f08}, where we compare the BH mass with AGN luminosity in each of the four Sy1 subgroups. In this figure the diagonals correspond to different Eddington ratios:
\begin{equation}\label{eq05}
{\rm N}_{Edd} \propto \frac{ {\rm L}_{AGN} }{ {\rm M}_{BH}} \propto \frac{ \eta\ \dot{\rm m}_{acc}}{ {\rm M}_{BH}}
\end{equation}
where $\dot{\rm m}_{acc}$ is the mass accretion rate and $\eta$ the efficiency (the fraction of mass transformed into light). From the positions of the Sy1s it is clear that those with detected AGN winds have higher N$_{\rm Edd}$ than their respective counterparts without wind (confirmed statistically in Table~\ref{stat2}), and this is despite having similar BH masses (cf. Table~\ref{stat1}). Therefore, the interpretation of these differences based on Eq.~\ref{eq05} is unambiguous: it means that the Sy1s with detected AGN winds have higher accretion rates than the Sy1s without detected winds. Or, considering the results of stacking, the strength of the AGN wind increases with the level of accretion. Two questions that naturally come to mind, then, are what explains this higher accretion and what effect could these winds have or had on their hosts?

Examining Table~\ref{stat1} and Table~\ref{stat2}, there are four physical characteristics that seem to be related only to the spectral difference B vs. N: 1) in the Sy1B and Sy1Bw, the BHs are more massive than in the Sy1N and Sy1Nw, 2) their power-law indices, $\beta$, are also lower (more negative), implying they are emitting more intensely in the UV, 3) their host galaxies have a slightly earlier morphology, and 4) their SFRs are slightly lower. The simplest scenario to explain all these differences seems to point to different galaxy formation processes: 1- according to the relation M$_{\rm BH}-\sigma_{\star}$, we expect the most massive BHs to be found in galaxies with more massive bulges, 2- this implies higher accretion rates in the Sy1B and Sy1Bw (early-type spirals) than in the Sy1N and Sy1Nw (late-type spirals) \citep{Coziol2011,Calvi2018}, 3- this difference in morphology also implies higher astration rates, that is, higher rates of transformation of gas into stars to form the bulges \citep{TinsleyLarson1979,StruckMarcell1981,Sandage1986,Coziol1998}, finally, 4- this implies that more metals and dust are locked into stars, leaving in the early-type spirals Sy1B and Sy1Bw bulges and disks depleted in gas and dust, compared to in the later type spirals Sy1N and Sy1Nw  bulges and disks rich in gas and dust. This scenario would naturally explain why the Sy1B and Sy1Bw emit more UV photons and why they have lower SFRs than the Sy1N and Sy1Nw (there is possibly also a direct connection with the formation of obscuring torus).

However, the general scenario above, based on galaxy formation process, does not explain why we find the same faction of detected AGN winds in the Sy1B and Sy1N. The key to understand the wind in these galaxies seems to be connected with the higher accretion rates and higher SFRs in the Sy1Bw and Sy1Nw. These observations imply more gas is falling onto the BH in these galaxies, and a good fraction of this ``free'' gas is sufficiently cold and dense to form more stars. Note that we can eliminate shocks due to AGN winds in the NLR, since we found in Figure~\ref{f3}c the intensity line ratios to be consistent with photoionization, and found no difference between the Sy1Nw and Sy1N. This suggests that the only mandatory condition to form an AGN wind could be a sufficiently large reservoir of gas available, which is consistent with the bias noted by \citet{Cicone2018}. Therefore, independently from how the galaxy formed, and thus of the final morphological type of the host galaxy, any ``surplus'' amount of gas reaching the BH might be sufficient to trigger an AGN wind. What is intriguing in this conclusion is that the physical conditions for the gas reaching the BH would also need to correspond to the physical conditions favoring star formation, which might be difficult to understand without a star formation triggering mechanism like an AGN wind feedback.

\begin{deluxetable*}{lcccccc}
\tablecaption{Spearman's correlation matrix for the whole sample of Sy1s \label{stat4}}
\tablewidth{0pt}
\tablehead{ \colhead{\textbf{Corr. Coeff.}}&\colhead{z}&\colhead{M$_{BH}$}&\colhead{L$_{\rm AGN}$}&\colhead{N$_{Edd}$}&\colhead{SFR}&\colhead{T}
} 
\startdata
M$_{\rm BH}$		&	\textbf{0.538}	&								&								&				&					&				   \\
L$_{\rm AGN}$	&	\textbf{0.849}	&	\textbf{0.654}	&								&				&			   	 	&					\\
N$_{\rm Edd}$	&	0.278					&	-0.370	 				&	0.273					&				&					&					\\
SFR						&	\textit{0.059}		&	\textit{-0.051}	&	\textit{0.085}		& 0.167	&					&				    \\
T							&	0.279					&	\textit{0.081}		&	0.223					&	0.152	&	0.339		&	 				\\
$\beta$				&	\textbf{-0.565} 	&	-0.334 					&	\textbf{-0.654} 	&	-0.274	&	-0.114	 	&	-0.426		 \\
\hline 
\textbf{P $\alpha = 0.05$}					&			z			&	M$_{BH}$	&	L$_{\rm AGN}$	&		N$_{Edd}$		&		SFR	&		T			\\
\hline
M$_{\rm BH}$			&	$< 0.0001$		&							&							&							&										&										\\
L$_{\rm AGN}$		&	$< 0.0001$		&	$< 0.0001$		&							&							&										&										\\
N$_{\rm Edd}$		&	$< 0.0001$		&	$< 0.0001$		&	$< 0.0001$		&							&										&										\\
SFR							&	0.0002				&	0.0015				&	$< 0.0001$		&	$< 0.0001$		&										&										\\
T								&	$< 0.0001$		&	$< 0.0001$		&	$< 0.0001$		&	$< 0.0001$		&	$< 0.0001$					&										\\
$\beta$					&	$< 0.0001$		&	$< 0.0001$		&	$< 0.0001$		&	$< 0.0001$		&	$< 0.0001$					&	$< 0.0001$					\\
\enddata			
\tablecomments {M$_{BH}$, L$_{\rm AGN}$, N$_{Edd}$ and SFR values are in dex (as in Table~\ref{stat1}); Strongest correlations ($rs \ge 0.5$) are in bold, weakest ($rs \le 0.1$) in italic.}
\end{deluxetable*}

\begin{deluxetable*}{lccccccc}
\tablecaption{Spearman's correlation matrix for Sy1 with resolved outflows \label{stat5}}
\tablewidth{0pt}
\tablehead{ \colhead{\textbf{Corr. Coeff.}}&\colhead{z}&\colhead{M$_{BH}$}&\colhead{L$_{\rm AGN}$}&\colhead{N$_{Edd}$}&\colhead{SFR}&\colhead{T} & $\beta$ 
} 
\startdata
M$_{\rm BH}$						&	\textbf{0.517}	&								&								&								&									&					& \\
L$_{\rm AGN}$					&	\textbf{0.837}	&	\textbf{0.627}	&								&								&									&					& \\
N$_{\rm  Edd}$					&	0.330					&	-0.445					&	0.355					&								&									&					& \\
SFR										&	\textit{0.072}		&	\textit{-0.057} &	0.108					&	0.216					&									&					& \\
T											&	0.324					&	\textit{0.087}		&	0.261					&	0.212					&	0.327						&					& \\
$\beta$								&	\textbf{-0.550} 	&	-0.317					&	\textbf{-0.666}	&	-0.362					&	-0.133						&	-0.443		& \\
V$_{\rm max}$ 					&	\textit{0.090}		&	0.165					&	0.135					&	\textit{-0.060}	&	\textit{-0.057}		&	-0.116		& 0.046\\
\hline 
\textbf{P $\alpha = 0.05$}	&			z					&	M$_{BH}$			&	L$_{\rm AGN}$		&		N$_{Edd}$		&		SFR					&		T						&  $\beta$\\
\hline
M$_{\rm BH}$					&	$< 0.0001$		&							&							&							&						&								& \\
L$_{\rm AGN}$				&	$< 0.0001$		&	$< 0.0001$		&							&							&						&								& \\
N$_{\rm Edd}$				&	$< 0.0001$		&	$< 0.0001$		&	$< 0.0001$		&							&						&								& \\
SFR									&	0.0057				&	0.0287				&	$< 0.0001$		&	$< 0.0001$		&						&								& \\
T										&	$< 0.0001$		&	0.0009				&	$< 0.0001$		&	$< 0.0001$		&	$< 0.0001$	&								& \\
$\beta$							&	$< 0.0001$		&	$< 0.0001$		&	$< 0.0001$		&	$< 0.0001$		&	$< 0.0001$	&	$< 0.0001$			&	\\
V$_{\rm max}$    			& 0.0006				& $< 0.0001$		& $< 0.0001$		& 0.0210				& 0.0287 		& $< 0.0001$			&	0.0764 \\
\enddata			
\tablecomments {M$_{BH}$, L$_{\rm AGN}$, N$_{Edd}$ and SFR values are in dex (as in Table~\ref{stat1}); Strongest correlations ($rs \ge 0.5$) are in bold, weakest ($rs \le 0.1$) in italic. Note that the correlation between V$_{max}$ and $\beta$ is not recognized as significant.}
\end{deluxetable*}

The second question, then, is what could be or could have been the effects of the AGN winds we observe on their host galaxies? Considering the stacking result, our analysis supports the idea that AGN winds are ubiquitous in Sy1s, which suggests this is an intrinsic aspect of the AGN phenomenon. But if this is so, then the effects expected of these winds must also be common. However, from our data we cannot distinguish what these effects could be. The only evidence we found is a weak trend for the  Sy1s with winds to have broad Balmer lines with smaller FWHMs than in Sy1s without wind, which suggests winds could have affected their BLRs (see Section~\ref{SS3a}). But that would be a local effect, while on a larger scale we found no difference between the Sy1N and Sy1Nw in terms of excitation in the NLRs. In order to get more information on this wind feedback problem, we computed a non parametric Spearman correlation matrix, which estimates the level of correlation for each pair of variables, without regard for the other variables. The Spearman correlation coefficient, $rs$, ranges from $-1$ to $+1$ (anticorrelation/perfect correlation), with $rs  =  0$ meaning no correlation. The accompanying matrix contains the p-values (for $\alpha = 0.05$), which when small means that we can reject the idea the correlation we observed are due to random sampling.

In Table~\ref{stat4} we present first the Spearman correlation and P-values matrices for the whole sample of 3,896 Sy1s. Because the matrices are symmetric, we present, for clarity sake, only the lower triangular parts. In Table~\ref{stat4} the strongest correlations are shown in bold, while the weakest ones are shown in italic. Based on the P-values, all the correlations are statistically significant. However, the strength of these correlations are not equal. The strongest correlations are also the most physically obvious, positive between M$_{BH}$ and L$_{AGN}$, and negative for these same two parameters with $\beta$, the index becoming more negative as the mass and luminosity of the AGN increase. The other strong correlations are with the redshift, showing rapid increase of BH mass and luminosity with the redshift (consistent with an increase of AGN activity at high redshift), which also explains the strong anticorreation with $\beta$. The correlation of the Eddington ratio, N$_{Edd}$, is also as expected based on its definition (Eq.~\ref{eq05}), increasing with the luminosity and decreasing with the BH mass. The fact that both of these parameters increase equally rapidly at high redshift might also explain the lower correlation of N$_{Edd}$ with $z$.

But more interesting are the results for the two parameters related to the hosts, the morphology type, T, and the SFR. What is remarkable is that their correlations with the parameters related to the AGN activity are significantly lower. As the morphology change with the redshift toward later-types, the BH mass almost does not vary, while both the AGN luminosity and Eddington ratio increase. This is consistent with the observational bias expected on the morphology related to an increase of AGN activity at high redshift. In fact, the only obvious physical correlation for T is the increase of SFR as the host galaxies change into later type spirals.  As for the SFR, the only other stronger correlations is with the Eddington ratio, consistent with what we observed before, implying that the SFR increases as the AGN activity increases. Note that this correlation could also explain the correlation of the SFR with $\beta$ and AGN luminosity. In general, therefore, there does not seem to be a a strong correlations between the SFR and AGN characteristics, or between the AGN activity and the morphological types (except for the bias).

Now that we have the general behavior, we can look what happens in those Sy1s with AGN winds. In Table~\ref{stat5} we computed the Spearman correlation matrix  adding V$_{max}$ as a supplementary parameter. The matrices shows mostly the same correlations as for the whole sample, with slightly higher coefficients, reinforcing the correlations of SFR with L$_{AGN}$ and N$_{Edd}$. This is consistent with the trend that Sy1s with winds show at the same time higher accretion rates and higher SFRs. As for V$_{max}$, the strongest positive correlations are with the BH mass and luminosity, consistent with what we expect of AGN winds. There is also a strong negative correlation with the morphology, the velocity decreasing in late-type spirals. This last correlation is consistent with what we found before,  V$_{max}$ being lower in the Sy1Nw, since Sy1Nw are more numerous in late-type spirals. The correlations of V$_{max}$ are much weaker and negative with N$_{Edd}$ and the SFR, and in fact there is no correlation (the only correlation rejected by the p-value) with the power-law index. 

These last results seem to confirm that the outflows we observed are AGN winds, and that there are no direct physical connection with what causes these winds and the SFRs in their hosts. The only significant anticorrelation of V$_{max}$ with T, which suggest that AGN winds become stronger in early-type galaxies, might be the only evidence that could suggest a link between AGN winds and the bulges. However, as we have mentioned in Section~\ref{SS2c}, the difference in velocity between the SyBw and Sy1Nw could be easily explained in term of winds by a difference of NLRs, which make this characteristic of the wind the product of a difference in morphology and not a cause of this difference in morphology. The only direct effect of the winds we observe might have been local, affecting the BLRs, which would explain why the Balmer lines in those Sy1s with winds tend to have smaller FWHMs, and this is independent from the spectral type (B versus N) and morphology.

Our observations, in particular, show no evidence favoring the quenching of star formation by AGN winds. At least not for the winds we observe, but then in what conditions would the winds, being ubiquitous and possibly intrinsic to the AGN phenomenon, be important? In \citet{Bait2017}, the authors concluded that, ``...the growth of the bulge plays an important role in quenching'', and ``morphology most strongly correlates with sSFR, independent of the environment...". Assuming AGN winds are not only ubiquitous, but transient and recurrent \citep[e.g.,][]{KingPounds2015}, could AGN winds play an active role in forming the bulge of the Sy1s in our sample? However, in Section~\ref{SS3c} we have determined that, assuming the mass of the bulge is 1000 times higher than the mass of their BHs, the sSFR of the Sy1s would be typical of early-type spirals (S0s to Sa/Sb) in the Green valley \citep[cf. Figure~8 in][]{Bait2017} far from the quenching region. There is consequently no evidence the SFRs we observe in the Sy1s are peculiar for the morphology and mass of their hosts. Thus if AGN winds played a role in the formation of these bulges, their effect would be today undistinguishable from the normal process of galaxy formation.

But, then, what about the effects of the wind we observe now? Could they have been delayed, as many authors suggested? The winds that we observe, consequently, would have formed only recently, and would not have had enough time to interfere with the SF of their hosts. However, how long could this delay last? Note that our sample already covers a large range in redshift, up to $z \sim 0.4$, which corresponds cosmologically to a 4 Gyrs look-back time, and surely after such a long time one should have expected feedback evidence to appear in sufficiently large sample. Or could it be that the feedback happens when the galaxy is already out of its AGN phase, explaining why no such case appears in our sample? But then what form would this post-AGN phase take and how would this transition happen? Would it be gradual or sudden? And how would that phase be compared to other AGN types, the Sy2s, LINERs, or TOs, which also have outflows \citep[e.g.,][]{Woo2016}, but significantly different AGN and star formation characteristics \citep[e.g.,][]{Torres-Papaqui2012, Torres-Papaqui2013}?.

Or maybe the effect of AGN winds is more direct (local). We already had a suggestion of that possibly happening in the BLRs (cf. Section~\ref{SS3a}). However, since the winds are ubiquitous, such effect would also need to be general, related to a common phenomenon that is obvious to observe (possibly something that we already observed). Interestingly, there is one well known phenomenon, which, although characteristic of AGN, is still unexplained, which is the fact that most AGN are radio quiet. In \citet{Coziol2017} it was suggested that AGN become radio-loud only when the accretion process in their galaxies becomes chaotic, which, on the other hand, they also demonstrated is a rare event, explaining why most AGN are radio quiet. However, why this is a rare event was not explained. Could AGN winds have something to do with this fact? For example, by ejecting a huge quantity of gas out of the central region, these winds could act as a natural mechanism to regulate the accretion rate, possibly impeding the process to become chaotic. This would make AGN winds not only ubiquitous, but intrinsic to the accretion process, in good agreement with our observations. What role AGN winds could play in radio galaxies, therefore, is an open question that needs to be investigated further.

\section{Conclusions}\label{S5}

The most significant result of our study is the confirmation that outflows are ubiquitous in Sy1s. We also found clear evidence that these outflows could be radiatively launched and that they are related to higher rates in gas accretion, consistent with AGN winds. This suggests that AGN winds are not only a common aspect of the AGN phenomenon, but most probably are intrinsic to the accretion process; they happen each time an extra amount of gas finds its way to the BH at the center  of the galaxies. 

What is not clear, however, is what could be the consequence of these winds? Having determined the SFRs and morphology of the host galaxies, we have found these parameters to be only weakly correlated with the parameters related to the source of the winds (BH mass, AGN luminosity and Eddington ratios). We also found that Sy1s with detected winds have higher SFRs than those without winds, which contradicts the original quenching hypothesis. Furthermore, we found that their specific SFRs are typical of early-type spiral galaxies (S0s to Sa/Sb) in the green valley, far from the quenching regime, which suggests that the Sy1 host galaxies are following a normal evolutionary path for their morphology and mass. 

Other interesting observations related to the winds are: 1) the  maximum velocity of the wind is higher in the Sy1Bw than in the Sy1Nw, which is consistent with denser NLRs in the Sy1Nw due to their later-type morphology, 2) Sy1s with wind, irrespective of their morphology, show a trend to have broad Balmer lines with smaller FWHMs than those without wind, which could be interpreted as a local effect of the wind in modifying the structures of the BLRs. 

Consequently, we propose that either the AGN winds in the Sy1s are recent occurrences, related to recurrent AGN events, and the winds, then, are too young to have had an observable influence on their galaxies, this happening only in post-AGN, or the feedback effects are mostly local, modifying the BLR and possibly regulating the accretion process itself.

\acknowledgments

The authors thank an anonymous referee for comments and suggestions that help us improve the clarity of our paper. They also want to thank Heinz Andernach for reading a first draft of this article and for his judicious comments. J. P. T.-P. acknowledges a grant support from DAIP-UGto (0173/19). R. C. also wants to thank Dr. Cindia Reyes (MD.) and the Drs. and nurses staff of the Centro Medico La Presa, in Guanajuato, for their excellent care in a time of needs during a difficult period (the COVID-19 pandemy), allowing him to continue working on this fascinating subject. The SDSS is managed by the Astrophysical Research Consortium (ARC) for the Participating Institutions. The Participating Institutions are: the American Museum of Natural History, Astrophysical Institute Potsdam, University of Basel, University of Cambridge (Cambridge University), CaseWestern Reserve University, the University of Chicago, the Fermi National Accelerator Laboratory (Fermilab), the Institute for Advanced Study, the Japan Participation Group, the Johns Hopkins University, the Joint Institute for Nuclear Astrophysics, the Kavli Institute for Particle Astrophysics and Cosmology, the Korean Scientist Group, the Los Alamos National Laboratory, the Max-Planck-Institute for Astronomy (MPIA), the Max-Planck-Institute for Astrophysics (MPA), the New Mexico State University, the Ohio State University, the University of Pittsburgh, University of Portsmouth, Princeton University, the United States Naval Observatory, and the University of Washington. This publication also makes use of data products from the Wide-field Infrared Survey Explorer, which is a joint project of the University of California, Los Angeles, and the Jet Propulsion Laboratory/California Institute of Technology, funded by the National Aeronautics and Space Administration and of the cross-match service provided by CDS, Strasbourg.


\bibliography{JPTP_etal2020V3}{}
\bibliographystyle{aasjournal}



\end{document}